\documentclass[aps,epsf,amsmath,superscriptaddress,citeautoscript,twocolumn,
showpacs,floatfix]{revtex4-2}
\usepackage{bm}
\usepackage{amssymb}
\usepackage{graphicx}
\usepackage{amsmath, amsthm, amssymb, graphicx}
\usepackage[usenames]{color}
\graphicspath{{figures/}} 
\usepackage{ulem}

\usepackage{amsmath,amsthm,amsfonts,amscd,amssymb}
\usepackage{eucal} 	 	
\usepackage{verbatim}      	
\usepackage{makeidx}       	
\usepackage{braket}

\usepackage{natbib}
\usepackage{bibentry}
\usepackage{hyperref}

\usepackage{mathrsfs}
\usepackage{graphicx}
\usepackage{dcolumn}
\usepackage{bm}

\begin{document}
	
	\preprint{}
	
	\title{Vacuum Tunneling of Vortices in 2-Dimensional $^{4}$He Superfluid Films}

\author{M.J. Desrochers}
	\affiliation{Physics and Astronomy, University of British Columbia, 6224 Agricultural
Rd. Vancouver, B.C., Canada, V6T 1Z1}
\author{D.J.J. Marchand}
\affiliation{1Qubit, 1950 Rue Roy
Sherbrooke, Quebec, Canada J1K 1B7}
\affiliation{Pacific Institute of Theoretical Physics, Univ. of British Columbia, 6224
Agricultural Rd. Vancouver, B.C., Canada, V6T 1Z1}
\author{P.C.E. Stamp}
\affiliation{Physics and Astronomy, University of British Columbia, 6224 Agricultural
Rd. Vancouver, B.C., Canada, V6T 1Z1}
\affiliation{Pacific Institute of Theoretical Physics, Univ. of British Columbia, 6224
Agricultural Rd. Vancouver, B.C., Canada, V6T 1Z1}

\begin{abstract}
At low temperature $T$ we expect vacuum tunneling processes to occur in superfluid
$^4$He films. We distinguish between extrinsic processes, in which single vortices
nucleate by tunneling off boundaries in the system, and intrinsic processes, in which
vortex/anti-vortex pairs quantum nucleate far from boundaries - this latter process is quite new. It is crucial to incorporate the
varying effective mass of the vortex in tunneling calculations. One intrinsic process
is the superfluid analogue of the Schwinger mechanism in quantum field theory; here
it appears as a quantum phase transition at $T=0$, driven by an external supercurrent.
We calculate the tunneling rate for these various vacuum tunneling processes, and describe a means of testing the predictions using a specific ``vortex counting" experiment.
\end{abstract}

\maketitle

Vacuum tunneling is an exotic process in quantum field theory, famously exemplified by
the``Schwinger process", in which electron-positron pairs spontaneously form out of the
vacuum in a very strong electric field \cite{schwinger}; in principle the same thing should occur in strong fields near sharp surfaces. These processes have never been seen
experimentally - the required fields ($\sim 10^{18} \, V/m$) are far too strong to be generated or sustained by any laboratory material. However in superfluids, vacuum tunneling is posited to occur by the quantum nucleation of vortices \citep{reviews,3d-expt,donnelly,sonin}. Superfluid $^4$He is distinguished by its simplicity, the large size of the roton gap, and the extraordinary purity with which it can be prepared. At low $T$ it then acts as an almost perfect quantum many-body vacuum.

We will here distinguish between two ``vacuum tunneling" processes in the superfluid:

(a) {\it Extrinsic tunneling processes}:  these are mediated by a boundary, and are thus
similar to the processes already studied in 3 dimensions \citep{reviews,3d-expt,donnelly,sonin}. They involve nucleation of single vortices at some point on the boundary, and their subsequent movement away from the boundary, driven by an external superflow. Experimental analysis is complicated because real boundaries are neither smooth nor simple.

(b) {\it Intrinsic tunneling processes}: these we discuss herein; they are in some ways much more interesting. They involve vortex/anti-vortex pair nucleation far from any boundary, in a background superflow. When the background superflow is uniform - which is easy to arrange experimentally - this process will be analogous to ``Schwinger'' vacuum tunnelling \cite{schwinger}, with the background superflow velocity $\boldsymbol{v}_{0}$ playing the role of the uniform electric field ${\bf E}$ in QED. However there is an important difference - although both the Schwinger process and the current superfluid vaacuum tunneling process involve fully quantum-mechanical vacua, the superfluid tunneling involves objects whose mass evolves during the tunneling process.

Until now the experimental evidence for {\it extrinsic} quantum nucleation of vortices has been entirely in 3-d superfluids, and is indirect; one infers it by the effect on the superflow past sharp boundaries in a 3-d superfluid, or around objects like ions moving through it. The vortex nucleation rate $\Gamma(T,\boldsymbol{v}_{s})$ flattens to a non-zero constant value below some crossover temperature $T_c$, when the superflow velocity $\boldsymbol{v}_{s}$ past a boundary in the superfluid (often the surface of some orifice) exceeds a critical value \citep{reviews,3d-expt,donnelly}. Above $T_c$, one sees thermally activated extrinsic nucleation. Interpretation of these experiments is hard because in 3 dimensions, vortex lines contort into a huge variety of complicated shapes, each with different tunneling and thermal activation rates; and the details depend on the structure of the surface from which the vortices nucleate, right down to atomic scales.

On the other hand, {\bf intrinsic} tunneling nucleation of vortices has never been seen in superfluids (theory has argued that {\it thermally activated} intrinsic processes should exist in 3 dimensional superfluids \cite{langer67}, but there is no clear evidence for this). When the background superflow is uniform, we will show that an intrinsic vortex tunneling process should be analogous to ``Schwinger'' vacuum tunnelling, with the background superflow velocity $\boldsymbol{v}_{0}$ playing the role of the uniform electric field ${\bf E}$. We argue that experiments looking for this should allow clean and unambiguous comparison with theoretical predictions.

In the present paper we discuss both extrinsic and intrinsic nucleation processes for 2-d superfluid films - until now nearly all theoretical discussions \cite{donnelly,sonin}  and experiments \cite{reviews,3d-expt} on vortex tunneling have been for 3-dimensional superfluids (but see ref. \cite{arovas08,ypsach}). 2-d superfluid films offer two key advantages:

(i) the geometry is much simpler, so that quantitative theory can be done, and compared unambiguously with experiment;

(ii) the tunneling rates should be much higher than in 3 dimensions.

As a result of the theory we give here, we find that both intrinsic and extrinsic vacuum tunneling of vortices should be observable in superfluid films. We will also see that the intrinsic processes can be discussed as a new kind of quantum phase transition (QPT), in which the coupling driving the transition is just the superfluid velocity
$\boldsymbol{v}_{s}$, where $\boldsymbol{v}_{s}$ is the local supercurrent velocity (for films, the normal fluid is locked to the substrate; henceforth we will work in the substrate frame of reference, so that the normal fluid velocity $\boldsymbol{v}_{n} = 0$). We will also see that one can think of this QPT as a continuation to $T=0$ of the well known Kosterlitz-Thouless (KT)
finite temperature phase transition \citep{thouless1,berezinskii,nelsonkosterlitz}.

In the course of doing this analysis one finds that all previous analyses of vortex nucleation in superfluids have suffered from a key flaw - the vortex effective mass has been treated as a constant, when in fact it varies radically as the vortex moves. In general the mass increases rapidly as the vortex moves away from a surface (for extrinsic nucleation), or as vortex/anti-vortex pairs move apart from each other (for intrinsic nucleation). Amongst many other things, this explains why vortex nucleation should be easier in 2 dimensions - the vortex mass increases less rapidly than in 3 dimensions.

One can think of vortex tunneling as a kind of macroscopic quantum tunneling (MQT) process, analogous to that in superconductors \citep{caldeiraleggett}, or in Ising ferromagnets \citep{LiHo24}. However, just as in those systems, the ``effective mass" of the tunneling entity is not macroscopic during the ``under barrier" tunneling process itself, but only after the entity emerges from under the barrier. Moreover, previous work on MQT has looked at the analogue of extrinsic tunneling - for example, tunneling in SQUIDs has looked at the tunneling of pre-existing fluxons. We are not aware of any experiments on \underline{intrinsic} vacuum tunneling processes (in which the tunneling objects are created out of the vacuum itself), in any laboratory system.

\vspace{4mm}


{\bf (i) VORTICES IN SUPERFLUID FILMS}: Superfluidity is found in $^4$He films when the
film thickness exceeds a couple of atomic layers \cite{chan74,hallock21}. The vortex
structure in a film is quite complex, and becomes more so as the vortex approaches a
boundary. Even in the absence of vortex-phonon interactions (which cause their own
complications \citep{kozikproksvis,thompsonstamp,forsachwool}), it is known that outside
the very small vortex core, of radius $\xi_0 \sim 1.7 \times 10^{-10}$m, there can, for thicker films, be a ``dimple''
structure, where centrifugal force lowers the superfluid surface over a length scale
$\sim$ 10-30 nm, depending on film thickness
\citep{harveyfetter,vittocolemeinke,rayfield}.

However, almost all the vortex energy is locked up in the long-range part of the superflow
field, allowing us to give accurate results for the vortex energetics. It also permits a
long wavelength theory for the vortex dynamics, valid for lengthscales $\gg \xi_0$, and
for energy scales $\ll \Lambda_{0}c_{0}^{2} \equiv \hbar c_o/\xi_0$, where $m_0$ is the
$^4$He atomic mass and $c_0$ the sound velocity. For $^4$He films, $\Lambda_0/\hbar \sim
150$ GHz.

To describe the vortex dynamics, we start from the work of Thompson and Stamp
\citep{thompsonstamp} who showed that
if one starts from the standard long-wavelength Bose superfluid action
\citep{faddeevpopov}, written in terms of phase fluctuations
$\phi\left(\boldsymbol{r},t\right)$
and density fluctuations $\eta\left(\boldsymbol{r},t\right)$ around
the moving vortex (which is centered at $\boldsymbol{R}_{V}$), then
one can write the vortex effective action in the form \cite{thompsonstamp}
\begin{align}
& S_{\text{eff}}^{V} \left[ \boldsymbol{R}_{V},\boldsymbol{\dot{R}}_{V}\right] =\int dt \;
\left(\tfrac{1}{2}M_{V}^o\dot{R}_{V}^{2} - {\cal T}(\boldsymbol{R}_{V})\, \right) \nonumber \\
&\qquad\qquad +  \int dt \left[
\boldsymbol{\dot{R}}_{V}\cdot\boldsymbol{A}_{V}\left(\boldsymbol{R}_{V},t\right)
- V_{V}\left(\boldsymbol{R}_{V},t\right) \right]
 \label{eq:Vortex Effective Action}
\end{align}
where $M_{V}$ is the vortex hydrodynamic mass, ${\cal T}(\boldsymbol{R}_{V})$ is a vortex potential energy, discussed below,
and where $\boldsymbol{A}_{V}\left(\boldsymbol{R}_{V},t\right)$ is the sum of a Magnus term ${\bf A}_0$ and a quasiparticle current \uline{pair}
field ${\bf J}(t)$, and is given by
\begin{equation}
\boldsymbol{A}_{V} = {\bf A}_0 + {\bf A}_{QP} \;\;=\;\;  \tfrac{1}{2}\left[\rho_{s}
\left(\boldsymbol{R}_{V}\times\boldsymbol{\kappa}\right)+{\bf
J}(t)\right]
 \label{eq:Quasiparticle Pair Field}
\end{equation}

Here $\rho_s$ is the superfluid mass density per unit area, ${\bf J}\left(t\right)=\int
d^{3}r\,{\bf j} ({\bf r},t)$
is the pair current (describing pairs of quasiparticles),
with pair current density
\begin{equation}
{\bf
j}\left(\boldsymbol{r},t\right)=\frac{\hbar}{m}\left(\eta\nabla\phi-\phi\nabla\eta\right)\label{eq:Quasiparticle
Pair Current}
\end{equation}
and $\boldsymbol{\kappa}= \pm (\hbar/m)\hat{\bf z}$ (for a vortex or anti-vortex). The Magnus term $\boldsymbol{A}_{0} = \frac{1}{2}\rho_{s}\left(\boldsymbol{R}_{V}\times\boldsymbol{\kappa}\right)$ causes the vortex to move perpendicular to the local flow field.

The potential $V_{V}\left(\boldsymbol{R}_{V},t\right)$ describes
vortex-quasiparticle interactions, for long-wavelength quasiparticles
interacting with the vortex superflow field; one has
\begin{equation}
V_{V}\left(\boldsymbol{R}_{V},t\right)=\frac{\hbar}{2m}\int
d^{2}r\,\boldsymbol{j}\left(\boldsymbol{r},t\right)\cdot\nabla\Phi_{V}\left(\boldsymbol{r}-\boldsymbol{R}_{V}\left(t\right)\right)
 \label{eq:Vortex-Quasiparticle Potential}
\end{equation}
In this equation we have defined $\Phi_V\left(\boldsymbol{r}-\boldsymbol{R}_{V}\right)$ as
the part of the total superfluid phase $\Phi\left(\boldsymbol{r},t\right)$
which comes from the vortex (so that $\Phi=\Phi_{V}+\phi$).
All the vortex-quasiparticle interactions in eqtns.(\ref{eq:Vortex Effective
Action})-(\ref{eq:Vortex-Quasiparticle Potential}) involve pairs of quasiparticles, not
single
quasiparticles \citep{thompsonstamp,QPpair}; in this we differ from other work on 2-d vortex dynamics \citep{arovas97}.

The effective mass of a superfluid vortex, even just the `bare"
hydrodynamic mass $M_V^o$, is a subtle quantity. The coupling of the vortex to quasiparticles like phonons and
rotons makes it even more so, because this coupling not only causes dissipation and drag
\cite{thompsonstamp}, but also renormalizes the hydrodynamic mass via radiative
corrections \cite{frousteycoxstamp}, and is linked to the frequency dependence of the effective mass (see also refs. \citep{arovas08,arovas97}. However, one can show that these corrections are small in $^4$He superfluid films \cite{frousteycoxstamp,supplements}, and we will ignore them here.

Arguments going back to Suhl, Duan, and others \citep{suhl,duan,popov,baym,niuaothouless} show that for a {\it single isolated vortex}, the hydrodynamic vortex effective
mass $M_V = {\cal T}/c_{0}^{2}$, where
$c_{0}$ is the sound velocity, and the ``potential energy" ${\cal T}$ is just the hydrodynamic superflow energy associated with the vortex. However this result has to be modified for multiple vortex systems (either when we have vortex/anti-vortex pairs, or when boundaries induce image vortices). The hydrodynamic flow energy then depends on the vortex coordinate, according to
\begin{equation}
{\cal T}({\bf R}_V) =  \tfrac{1}{2} \int d^2r \; \rho_s({\bf r} - {\bf R}_V)\, v_S^2({\bf
r})
 \label{cal-T}
\end{equation}
where this potential energy automatically incorporates the Magnus term. Let us now consider how ${\cal T}({\bf R}_V)$ varies for two model systems where boundaries play a role.

\vspace{4mm}


{\bf (ii) EXTRINSIC NUCLEATION}: Consider first a vortex moving inside a circular
boundary of radius $R_{0}$, at a distance $R_{V}$ from
the center of the circle (see Fig. \ref{fig:Vortex-1}). The superfluid vortex energy, including the Magnus term, is then given by long-wavelength hydrodynamics as
\begin{equation}
{\cal T}({\bf R}_V) \;=\; \frac{\rho_{s}\kappa^{2}}{4\pi} \,
\ln\left(\frac{R_{0}^{2}-R_{v}^{2}}{R_{0}
\xi_0}\right)-\kappa\omega_{0}\rho_{s}\left(R_{0}^{2}-R_{V}^{2}\right)
 \label{eq:Vortex
Energy - Circular Boundary}
\end{equation}
where $\kappa=h/m_0$ is the circulation quantum, {$\xi_{0}$ is the vortex
core radius, and $\rho_{s}$ the areal superfluid density;
the circle (with any remaining entrained normal fluid) is then rotated at angular
velocity $\omega_{0}$.

Let us now consider the geometry
in Fig. $\text{\ref{fig:Vortex-1}}$, where the superflow velocity increases around a semi-circular protuberance
on the boundary, favouring quantum
and thermal nucleation at the surface of this protuberance. We assume $R_{0}\gg r_0$, the protuberance
radius. The problem then reduces to one with a supercurrent
$\boldsymbol{v}_{s}\left(\boldsymbol{r}\right)$
moving past the protuberance such that
$\boldsymbol{v}_{s}\left(\left|\boldsymbol{r}\right|\gg
r_0\right)\rightarrow\boldsymbol{v}_{0}$
(the ``background" supercurrent, far from the protuberance). In what follows we will
assume that $R_0$ is so large that it can be treated as infinite, so that on
scales $\sim r_0$, the circular boundary can be treated as rectilinear, with
 $\boldsymbol{v}_{0}$ parallel to this boundary.

One can then show (see supplementary information
\citep{supplements}) that the hydrodynamic superflow energy (again, including the Magnus term) due to a vortex at position ${\bf
R}_V = (R_V, \alpha)$
with respect to the protuberance centre (with $R_V > r_{0}$ and angle $\alpha$
with respect to $\boldsymbol{v}_{0}$), is given by  \citep{supplements,marchand}:
\begin{widetext}
\begin{eqnarray}
\mathcal{T}( {\bf R}_V)\;\;&=&\;\;\frac{\rho_{s}\kappa^{2}}{8\pi}\left\{
\ln\left(\frac{4R_{V}^{2}\sin^{2}\alpha}{\xi_{0}^{2}}\right)
- \ln\left[\left(\frac{R_{V}^{2}+r_{0}^{2}}{R_{V}^{2}-r_{0}^{2}}\right)^{2}\sin^{2}\alpha+\cos^{2}\alpha\right]  \right\}  \nonumber\\
&& \qquad\qquad\qquad - {\rho_s\kappa v_s^o \over 4} \left\{ \left(\frac{R_{V}^{2}- r_{0}^{2}}{R_{V}}\right)\sin\alpha \,+\,  {8 r_0 \over \pi} \tan^{-1} \left(\frac{r_{0}R_{V}\sin \alpha}
 {R_V^{2}-r_{\text{0}}^{2}}\right) \right\}  \;-\; {\pi \rho_s r_0^2 \over 4} (v_s^o)^2
  \label{eq:Energy-Bump}
\end{eqnarray}
\end{widetext}

A plot of this potential energy is shown in Fig. \ref{fig:BumpPotential}, and one sees that the tunneling off the protuberance will be through an energy barrier. As one increases $|\boldsymbol{v}_{0}|$, the potential barrier preventing vortex nucleation gradually decreases in height and width - at a very high ``critical" velocity $v_{c}$ the barrier disappears. We will see that in experiments, tunneling becomes appreciable at a much lower velocity.


\begin{figure}
\includegraphics[width=0.5\textwidth]{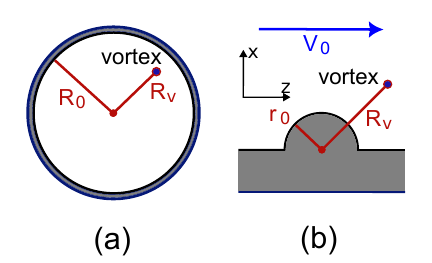}
\caption{Two geometries for extrinsic processes; we view the superfluid from above (looking down on the xy-plane. In (a) we have a 
vortex in a circular cylinder of radius $R_0$, with
the vortex at a distance $R_V$ from the centre. In (b) superflow moves at velocity
$\boldsymbol{v}_s$ along the $\hat{x}$ axis past a semicircular protuberance or ``bump" attached to a straight boundary with surface 
normal along the $\hat{y}$ axis. The vortex is at position
vector ${\bf R}_V$, at distance $R_V$ from the bump centre, and ${\bf R}_V$ has angle
$\alpha$ with respect to $\boldsymbol{v}_s$. }
\label{fig:Vortex-1}
\end{figure}


Consider then the tunnelling problem for this barrier. Vortex tunneling calculations
have been done before \citep{reviews,donnelly,sonin,marchand,volovik}, with varying results (for a more detailed comparison, see ref. \citep{supplements}). One reason
for this variety has been confusion over what effective mass to choose for the vortex
\citep{duan,popov,baym,thouanglin,niuaothouless}, and previous attempts to describe
tunnelling of vortices in superfluids have all assumed either a constant vortex mass (sometimes chosen as the effective mass of the vortex core), or a mass equal to zero. In reality, however, the dynamic effective mass of the vortex, $M_{V}\left({\bf R}_{V}\right)$ {\it varies with the vortex position}, both during tunnelling and after the vortex emerges outside of the barrier, and depends on the entire flow field associated with the vortex. In the case of extrinsic nucleation, the relation $M_V = {\cal T}/c_{0}^{2}$ found for an isolated vortex \citep{suhl,duan,popov,baym,niuaothouless} is then no longer valid - we now deal with multiple vortices.


\begin{figure}
\includegraphics[width=0.5\textwidth]{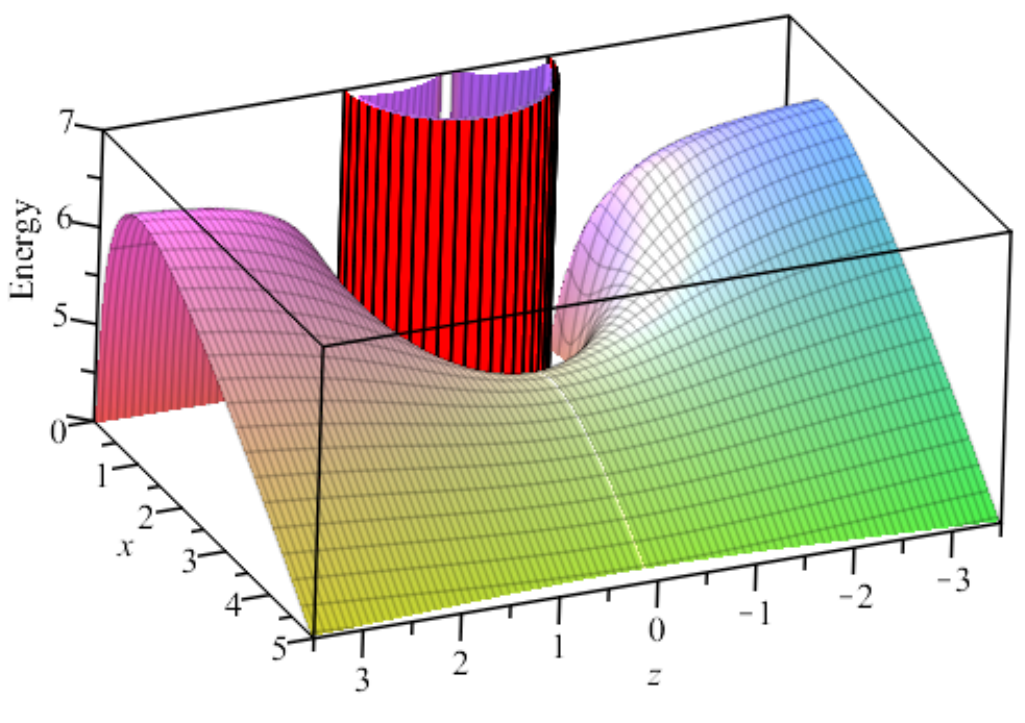}
\caption{A plot of the hydrodynamic superflow energy for the semicircular bump geometry of
Fig \ref{fig:Vortex-1}(b), where this bump is depicted in red and the energy surface in
blue-green. One can see that the tunneling probability is maximized along the trajectory
$\alpha=\frac{\pi}{2}$, shown as a white line in the figure.}
\label{fig:BumpPotential}
\end{figure}


However one can generalize the derivation given for an isolated vortex to this situation.

To describe extrinsic tunneling, we start from a path integral description of the vortex propagator, which we will write as
\begin{equation}
K_V(2,1) \;=\; \int {\cal D}{\bf P}_v \int {\cal D} {\bf R}_v \; e^{i S_V[{\bf P}_v, {\bf R}_v]/\hbar}
 \label{Kv-21}
\end{equation}
where here the vortex momentum is taken to be ${\bf P}_v = M_V \dot{\bf R}_v + {\bf A}_0$. In this expression we have dropped the functional integration over the quasiparticle variables, leaving only the Magnus term and the kinetic and potential terms. This is done because the "radiative corrections" to the effective vortex mass coming from the quasiparticles are found to be small \citep{frousteycoxstamp,supplements}, indicating that the corrections to the tunneling rate caused by quasiparticle dissipation should also be fairly small (we stress that this last argument needs to be substantiated by a detailed calculation \citep{supplements}).

In all the tunneling problems to be studied here, the Magnus term is actually parallel to the minimum action tunneling path, and so it can be absorbed directly into the potential term ${\cal T}({\bf R}_V)$.

Under these circumstances the path integral takes the form $K_V(2,1) \rightarrow
K_V^{(0)}(2,1)$, where
\begin{equation}
K^{(0)}_V(2,1) \;=\; \int {\cal D} {\bf R}_v e^{i S^{(0)}_V[{\bf R}_v, \dot{\bf R}_v ]/\hbar}
 \label{K0v-21}
\end{equation}
in which the action is now
\begin{equation}
S^{(0)}_V[{\bf R}_v, {\bf \dot{R}}_v ] \;=\; \int dt \;\left[
\frac{1}{2}M_{V}(\boldsymbol{R}_{V}) \dot{R}_{V}^{2} - {\cal T}(\boldsymbol{R}_{V}) \right]
 \label{S0-action}
\end{equation}

It is now straightforward to convert this to a WKB form for the tunneling rate, using standard instanton methods \cite{supplements}; we get
\begin{equation}
\Gamma_{\text{WKB}}=\Omega_{0}\exp\left\{ -\frac{2}{\hbar}\int_{a}^{b}d{\bf R}_V \,\left[2M_V ({\bf R}_V)\, {\cal T} ({\bf R}_V) \right] ^{\frac{1}{2}}\right\}
 \label{eq: WKB Tunneling Rate}
\end{equation}
where the effective potential though which the vortex tunnels is now ${\cal T}({\bf R}_V)$, and and $a$ and $b$ are the tunneling end-points. The position of these end-points, the form of ${\cal T}({\bf R}_V)$, and the related size of the prefactor $\Omega_0$, are governed by short-range processes at scales
$\sim \xi_0$, and energy scales $\sim \Lambda_0$. For superfluid $^4$He films we have a rough value for $\Omega_0
\sim \Lambda_0/\hbar \sim 150$ GHz, and this allows us to find the end-points numerically \cite{supplements}.

For the protuberance problem, ${\cal T}\left(\boldsymbol{R}_{V}\right)$ is given by (\ref{eq:Energy-Bump}), with a ``minimum action" WKB tunneling path along the straight line $\alpha=\frac{\pi}{2}$. Along this path $M_V (R_v) = (\rho_s \kappa^2/ 4\pi c_o^2) \,\ln \;| (R_v - r_0)/ \xi_o |$ for $(R_v - r_0) \ll r_o$ (ie., under the barrier), increasing steadily as the vortex moves away from the surface of the protuberance. This increase of the vortex mass increase is a general feature of any extrinsic vortex tunneling problem, and (see below) also occurs in intrinsic tunneling processes.

The integral under the tunnelling barrier expressed by (\ref{eq:Energy-Bump}) and visualized in Fig. \ref{fig:BumpPotential} admits neither an analytic solution, nor an easy expression for the tunnelling bounds $a,b$. We therefore evaluate numerically the tunneling action when $T=0$ (for details see the supplementary information \citep{supplements}).

Because the vortex mass is initially rather small, the vortex tunnelling
initially proceeds rapidly, then slows down under the barrier as the mass increases.

Let us emphasize what has been left out of this $T=0$ calculation.
We have ignored the dissipative coupling of the vortex to quasiparticles,
which operates even at $T=0$ through the couplings in $\boldsymbol{A}_{V}$
and $V_{V}$ in $\text{\eqref{eq:Vortex Effective Action}}$. Their effect on tunneling  can be analyzed in detail using a generalization of
Caldeira-Leggett theory \citep{caldeiraleggett}, but the results are rather complicated and will be addressed elsewhere \citep{desrochersstamp}. These effects are also crucial
for a correct evaluation of `over-barrier' thermally activated vortex nucleation.

Preliminary estimates suggest that corrections to the bare WKB result will be small,
simply because the corresponding corrections to the real time vortex equation of motion go as a high power of temperature \cite{thompsonstamp}, and because the corresponding radiative corrections to the effective mass are also small \citep{frousteycoxstamp}. This argument is similar to that used in Simon et al. \cite{LiHo24}, to neglect dissipative corrections to magnetic domain wall tunneling. In the case of 2-d vortices, the radiative corrections to the effective mass are $\sim O(m_o/M_V^o) \sim m_0 (a_o/L_z)/ ln |R/\xi_o|$, where $R$ is the lengthscale over which the vortex flow field extends, and $a_o \sim 2 \xi_0$ is the interparticle spacing. Even for very thin superflud films, this ratio is small and rapidly becomes smaller as $R$ increases (even the thinnest superfluid films \cite{chan74,hallock21} have $a_o/L_z < 0.4$).

\vspace{4mm}


{\bf (iii) INTRINSIC NUCLEATION}:  We now look at the more and novel and interesting process of intrinsic nucleation, in some region far from any boundary. The rest frame is now set by the substrate and any residual normal fluid locked to it. Suppose
we again have a uniform background superflow field with velocity
$\boldsymbol{v}_{s}$$\left(\boldsymbol{r}\right)\rightarrow\boldsymbol{v}_{0}$,
and ask - what will happen as we again increase $\boldsymbol{v}_{0}$? The
answer was given in the introduction: we get the spontaneous formation
of vortex/anti-vortex pairs, in a process precisely parallel to the
formation of electron-position pairs in the QED vacuum under a strong
electric field \citep{schwinger}. Using what we have just worked out above, it
is relatively simple to find the $T=0$ rate for this process.

We define vortex/anti-vortex coordinates $\left\{ \boldsymbol{R}_{j}\right\} $,
with $j\in\left\{ 1,2\right\}$. In a straightforward generalization of the analysis for a
single vortex, we can write a tunneling action for a pair of vortices \citep{supplements}.
Writing sum and difference coordinates
$\boldsymbol{Q}=\frac{1}{2}\left(\boldsymbol{R}_{1}+\boldsymbol{R}_{2}\right)$
and $\boldsymbol{r}_{12}=\left(\boldsymbol{R}_{1}-\boldsymbol{R}_{2}\right)$, one easily
evaluates ${\cal T}\left(\left\{\boldsymbol{R}_{j}\right\} \right)$
for a vortex/anti-vortex pair separated by distance $\boldsymbol{r}_{12}$.


\begin{figure}
\includegraphics[width=0.5\textwidth]{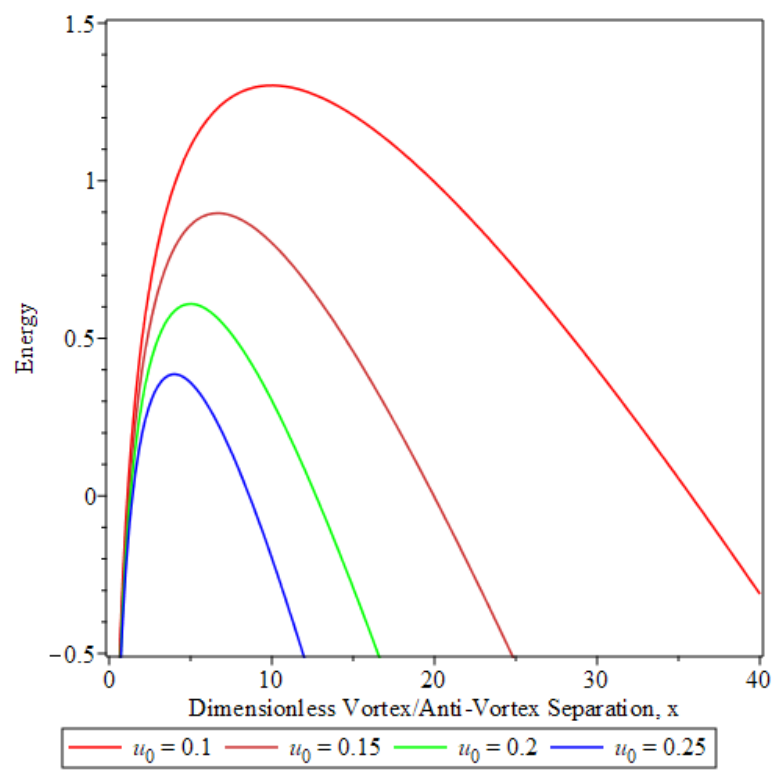}
\caption{A plot of the vortex/anti-vortex pair potential, in units of $E_0=\rho_s
\kappa^2/4\pi$ in \eqref{eq:Vortex-Anti-Vortex WKB action} for various values of the
dimensionless external superflow velocity $u_{0}=v_0/v_c= 2\pi xi_0 v_s/\kappa$ and
where the superflow critical velocity is $v_{c} = \kappa/2\pi xi_0$.  }
\label{fig:PairPotential}
\end{figure}


We get at $T=0$, the tunnelling action in the form given in (\ref{eq: WKB Tunneling Rate})
above, but now with
\begin{equation}
{\cal T}({\bf r}_{12}) \;=\;
\frac{\rho_{s}\kappa^{2}}{4\pi}\ln\left|\frac{r_{12}}{\xi_{0}}\right| -
\frac{1}{2}\rho_{s}
\kappa\boldsymbol{v}_{0}\cdot {\bf r}_{12}
  \label{eq:Vortex-Anti-Vortex WKB action}
\end{equation}
with the potential barrier shown in Fig. \ref{fig:PairPotential}, for a
path ${\bf r}_{12}$ along the path of minimum tunneling action, such that ${\bf r}_{12}$
is perpendicular to $\boldsymbol{v}_{0}$. Along this path, the effective mass  varies as $M_V({\bf r}_{12}) = (\rho_s \kappa^2/ 4\pi c_o^2) \ln |r_{12}/\xi_o|$, increasing as the vortices separate. This tunnelling potential has a maximum when the vortex/anti-vortex pair is separated by a distance $r_{c}=\kappa/4\pi v_{s}$.

The effect of the supercurrent is to try to pull the vortex/anti-vortex
pair apart with a constant force
$\frac{1}{2}\rho_{s}\kappa\boldsymbol{v}_{0}\cdot\hat{\boldsymbol{r}}_{12}$,
where $\hat{\boldsymbol{r}}_{12}= \boldsymbol{r}_{12}/\left|\boldsymbol{r}_{12}\right|$.
This force is resisted by the attractive logarithmic potential between
them, creating the potential barrier shown in the Figure. As
the external superflow velocity increases, the barrier decreases; again it eventually
disappears at a ``critical" velocity $v_c = \kappa/2 \pi \xi_0 \sim 60$ ms$^{-1}$. However, even for $|\boldsymbol{v}_{0}| \ll v_c$, one expects tunneling through the barrier at $T=0$, and thermal activation for even lower velocities at higher $T$.

This is precisely what we found for the case of tunneling nucleation of a single vortex off a surface. However here the physics is quite different - we now deal with a process of \underline{pure} vacuum tunneling, where vortex/anti-vortex pairs appear spontaneously in the bulk superfluid, in clear analogy with the Schwinger process
\citep{schwinger}. The use of extremely pure $^4$He should make this an ideal candidate for experiments on this phenomenon.

By a parallel analysis to that given above, we end up with a zero-$T$ WKB tunnelling rate for intrinsic tunneling which, like the extrinsic case, must be found numerically. It is
given in integral form by $\Gamma_{\text{WKB}} = \Omega_0 \exp [-\gamma(u_0)]$, where
\begin{equation}
\gamma(u_0) =\frac{\sqrt{2}\xi_{0}\rho_s \kappa^2}{2 \pi \hbar c_{0}} \psi_{ba} \;\sim\; 0.88 {L_z \over a_o} \psi_{ba}  
 \label{eq:Vortex/Anti-Vortex Pair Tunneling Rate}
\end{equation}
and the dimensionless factor $\psi_{ba}$ is given by
\begin{equation}
\psi_{ba} \;=\; \int_{x_{a}}^{x_{b}}dx\,\sqrt{\ln\left|x\right|\left(\ln\left|x\right|-u_{0}x\right)}
 \label{psi-ba}
\end{equation}

Here $u_{0}=2\pi\xi_{0}v_{0}/\kappa$ and $x=r_{12}/\xi_{0}$ are dimensionless variables
for the background superflow velocity and vortex/anti-vortex separation; and $x_a,x_b$ are provided by Lambert W functions. As before, $\Omega_{0}$ must be fixed
numerically. The beginning and end points for tunneling are known as well \citep{supplements},
\begin{eqnarray}
x_{a}\left(u_{0}\right) \; &=& \; -\; \text{LambertW}\left(0,-u\right) \nonumber \\
x_{b}\left(u_{0}\right) \;&=& \; -\; \text{LambertW}\left(-1,-u\right)
\end{eqnarray}
}

Again, this is a $T=0$ result; and it also ignores the dissipative coupling to quasiparticles. Again, we argue that dissipative corrections to this rate will not be large, for the same reasons as above. It is amusing to note that there must also be dissipative corrections to the classic Schwinger calculation of vacuum tunneling in QED, but these corrections will be very small, because the fine structure coupling constant $\alpha$ is so small.

The zero-$T$ tunneling rate $\Gamma_{WKB}(u_0)$ varies extremely rapidly with $u_0$ (for multiple plots see ref. \cite{supplements}). One finds a fairly sudden onset for $u_0 \rightarrow u_c^{exp} \sim 0.1$, for a single layer of superfluid; since $v_c \sim$ 60$\; ms^{-1}$, this indicates a rapid {\bf experimental} crossover to observable tunneling when the external superflow velocity $v_o \rightarrow v_c^{exp} \sim$ 6$ \; ms^{-1}$. This value will be lowered even further by the vortex-quasiparticle coupling, but raised by increasing the film thickness \citep{supplements,desrochersstamp}.

\vspace{4mm}


{\bf (iv) SUGGESTED EXPERIMENTAL TEST}: To the best of our knowledge, no experiments have yet looked for or found vortex tunneling in superfluid films. Yet the theory herein shows that very clean experiments ought to be possible, and that, because the effective mass of the vortex can be made small for thin films (considerably smaller than for 3-d superfluids), tunneling ought to be easily visible.  The theory
also suggests several different possible experiments.
The most obvious and practically feasible of these involves driving a superflow
past an obstacle. This obstacle could be semi-circular as in (\ref{eq:Energy-Bump})
above, or circular (by reflection symmetry, the result given above in
(\ref{eq:Energy-Bump}) still applies), or oval (for which the calculations given
above are easily adapted using a conformal transformation).

Alternatively one can attempt to drive a homogeneous superflow in
a region far from any boundary, in order to see the entirely novel process of intrinsic vacuum tunneling, to which the tunneling rate in eqtn.
(\ref{eq:Vortex/Anti-Vortex Pair Tunneling Rate}) applies.

To test the predictions given here, we suggest the following scheme.
Given that one is now able to observe, in principle \cite{ypsach}, the
appearance and subsequent dynamics of individual vortices in 2-dimensional
films of superfluid $^{4}$He, we can do a simple counting experiment. This would just mean
counting the total number
$n\left(t\right)$ of single vortices that nucleate off the obstacle
as a function of time $t$. Then $\dot{n}\left(t\right)=dn/dt$
is given by \citep{SQUID}
\begin{align}
\dot{n}\left(t\right) &
=\Gamma_{\text{WKB}}\left(\boldsymbol{v}_{s}\left(t\right)\right)\label{eq:Vortices
per unit time.}\\
 & =\Gamma_{\text{WKB}}\left(v_{s}\right)\exp\left\{
 -\int_{-\infty}^{t}dt^{\prime}\,\Gamma_{\text{WKB}}\left(v_{s}\left(t^{\prime}\right)\right)\right\}\nonumber
\end{align}
where we assume that we are also allowed to slowly change $v_{s}\left(t\right)$
with time. Now define the ``switching distribution'' $\mathcal{P}=\frac{dn}{dv_{s}}$;
we then have
\begin{align}
\mathcal{P}\left(v_{s}\right) &
=\left(\frac{dv_{s}}{dt}\right)^{-1}\Gamma_{\text{WKB}}\left(v_{s}\right)\label{eq:Switching
Distribution}\\
\times & \exp\left\{
-\int_{0}^{v_{s}}dv_{s}^{\prime}\,\left(dv_{s}^{\prime}/dt\right)^{-1}\Gamma_{\text{WKB}}\left(v_{s}^{\prime}\right)\right\}\nonumber
\end{align}

We can apply the same argumentation to study the nucleation of vortex/anti-vortex pairs
far from boundaries: the same result $\text{\eqref{eq:Switching Distribution}}$
applies.

Experiments can also plot the behaviour of
$\mathcal{P}\left(v_{s};T\right)$ at finite temperature $T$: one
expects a crossover to thermally activated nucleation, the details
of which need to be worked out \citep{desrochersstamp}. At $T=0$, one will see a crossover on increasing $v_o$, from no observable tunneling to extremely rapid transitions as one crosses through $v_c^{exp}$. At first, $dn/dt$ will be low enough to count individual vortices; but above $v_c^{exp}$ it will become extremely high, making it impossible to resolve the passage of individual vortices. In this case one can instead measure the superflow dissipation caused by the quasi-continuous vortex generation.

The principal problem confronting experiments on any extrinsic tunneling process are (i)
uncontrolled surface irregularities, which will make the tunneling rates and critical
velocities uncontrollable; (ii) the possible excitation of internal 3-d quasiparticle modes, including Kelvin modes (``Kelvons")
in the vortex, and quasiparticles perpendicular to the film plane; and (iii) other possible distortions of the vortex at the boundary. For {\it intrinsic} nucleation however, far from boundaries, we
only have to worry about Kelvin modes and 3-d quasiparticles. For a film of thickness $L_z$, the energy of the
lowest Kelvon is $\tilde{\omega}_K^0 \equiv \omega_K(k = \pi/L_z) = (\pi^2 \hbar/2 m_0
L_z^2) \ln (L_z/\pi \xi_o)$, and the lowest energy quasiparticles along $\hat{z}$ have energy $\omega_{3d}^0 = \hbar c_0 (\pi/L_z)$. For $L_z = 10 \xi_o$, one then finds $\tilde{\omega}_0 \sim 1K$, and $\omega_{3d}^0 \sim 0.6K$. For $L_z = 50 \xi_0$, we get $\tilde{\omega}_0 \sim 104 mK$, and $\omega_{3d}^0 \sim 120mK$. Thus quite low temperatures will be needed, unless one uses thin films \citep{KeZhou}.

\vspace{4mm}


{\bf (v) REMARKS and CONCLUSIONS}:  Our central results here are quantitative predictions
for the quantum nucleation of vortices in different superfluid film geometries. Two key
differences from all previous work are (i) the recognition that the effective mass of the
vortices varies continuously during the tunneling process, which radically alters the
predictions; and (ii) the predictions for intrinsic vacuum tunneling, which should be
quantitatively testable without the usual complications attending 3-dimensional extrinsic tunneling processes, provided we use fairly thin films and sufficiently low temperatures \citep{KeZhou}.

One can consider many other geometries apart from those discussed here, and mimic several
interesting models in high-energy physics and quantum gravity \citep{desrochersstamp}. For
example, we can envisage an experiment in which superfluid flows towards a circular
``drain" hole or ``sink". The superfluid accelerates towards the sink, and in a critical
circular region around the hole, vortex/anti-vortex pairs are
produced. One then finds that ${\cal T}({\bf r}_{12}, r)$ is given by an expression
similar to
(\ref{eq:Vortex-Anti-Vortex WKB action}), but now both $\rho_s(r)$ and
$\boldsymbol{v}_{0}(r)$
depend on the radial distance from the sink point. This experiment then mimics Hawking
radiation around a black hole, with the critical region mimicking the horizon. Such
analogues are quite interesting, since one cannot do experiments on the original. Note here that our superfluid analogue here is a genuine quantum vacuum, not some semiclassical approximation to it.

However one can also argue that, instead of just studying analogues (which may or may not accurately portray the ``real thing"), it is much more interesting to explore the new physics involved in a real life condensed matter vacuum tunneling process. Here we emphasize that for instrinsic tunneling nucleation, we are dealing at $T=0$ with a new kind of quantum phase transition (QPT), in which the coupling driving the QPT is now just the external flow field velocity $v_0$.

In a 3-dimensional system one can imagine continuing a finite-$T$ 1st-order transition,
driven by the nucleation of critical bubbles, down to $T=0$, where it will occur through
tunneling nucleation - such scenarios have been considered in both quantum cosmology
\cite{coleman} and in, eg., superfluid $^3$He \cite{bailin80}.
However in superfuid $^4$He films we have something rather different. At $T=0$ the
intrinsic transition is driven by the flow field $v_0$; but when $v_0$ is small or zero, a
transition occurs at finite $T$  by the usual Kosterlitz-Thouless vortex/anti-vortex
unbinding mechanism  \cite{thouless1,nelsonkosterlitz}. Thus there will be a continuous
evolution between the two kinds of transition as one increases $v_0$ in the range $0 < v_0
< v_c$. To analyze this will require a modification of KT theory, and should provide
interesting predictions for experiment.

Finally, we note that with a continuous tunneling generation of vortices or anti-vortex
pairs, in a ``quantum avalanche" process of the kind already seen in quantum Ising systems
\cite{LiHo24}, one will end up with a an interacting vortex gas in which a 2-dimensional
quantum turbulent state \cite{Qturb} can be generated \citep{desrochersstamp}.

\vspace{4mm}

{\bf ACKNOWLEDGEMENTS}: We thank C.Baker, W.Bowen, P. Dosanjh, D.Harvey, V. Milner, and K. Zhou for discussions. This work was supported by the National Science and Engineering Research Council in Canada.

\vspace{4mm}

\begin{widetext}

{\bf MATERIALS and METHODS}: This work is a combination of analytic and numerical theory; no experiments were performed. The supplementary information \cite{supplements} gives further details of the theory. We describe therein (i) how real superfluid films, with or without vortices, can be modelled, and how for thin films this can be reduced to a description in terms of meromorphic complex functions in bounded or unbounded domains; (ii) how the energetics and effective masses of vortices are calculated both in bounded films, including calculations for a vortex nucleating off a semicircular protuberance on a container wall, and for vortex/anti-vortex pair nucleation in an unbounded film; (iii) how vortices interact with quasiparticles, and why these interactions can be neglected, to good approximation, for tunneling calculations in realistic films; and (iv) how the theoretical modelling and calculation of vortex tunneling rates, for bounded and unbounded films, is done. The details of the numerical methods employed to calculate the tunneling rates are also given in the supplementary information, along with cross-checks of the results.

\end{widetext}



\begin{thebibliography}{99}

\bibitem{schwinger}    J.Schwinger, Phys.Rev. {\bf 82}, 664 (1951). Note that our results
    for superfluid films can be compared with a Schwinger-type calculation, but done in
    $2+1$ dimensions (where the Coulomb interaction is logarithmic).

\bibitem{reviews}    For reviews on vortex nucleation in 3-dimensional superfluids, see,
    eg., R.E. Packard, J.C. Davis, Physica B{\bf 197}, 315 (1994); E. Varoquaux, Comptes
    Rendus Physique {\bf 7}, 1101 (2006); E. Varoquaux, Rev.Mod.Phys. {\bf 87}, 803
    (2015); and references therein.

\bibitem{3d-expt}         Some early experiments on vortex tunneling in 3-dimensional
    $^4$He superfluid are described in R.M. Bowley, P.V.E. McClintock, F.E. Moss,
    G. Nancolas, P.C.E. Stamp, Phil. Trans. Roy. Soc. A{\bf 307}, 201-260 (1982); J. C. Davis,
    J. Steinhauer, K. Schwab, Yu. M. Mukharsky, A. Amar, Y. Sasaki, and R. E. Packard,
    Phys. Rev. Lett. {\bf 69}, 323 (1994); E. Varoquaux, O. Avenel, Phys. Rev. B {\bf 68},
    054515 (2003); later experiments are described in the reviews of ref. \cite{reviews}.

\bibitem{donnelly}    R.J. Donnelly, ``{\it Quantized Vortices in Helium II}" (Cambridge University Press, 1991), particularly Chapter 8.

\bibitem{sonin}       Eduoard B. Sonin, ``{\it Dynamics of Quantized Vortices in Superfluids}" (Cambridge Unversity Press, 2016), particularly Chapter 11.


\bibitem{langer67}    J.S. Langer, J.D. Reppy, Prog. Low Temp. Phys. VI, pp. 1-34
    (North-Holland, 1970)

\bibitem{arovas08}     D.P. Arovas, A. Auerbach, Phys. Rev. {\bf B}78, 094508 (2008)

\bibitem{ypsach}          Y.P. Sachkou et al., Science {\bf 366}, 1480 (2019)

\bibitem{thouless1}      J.M. Kosterlitz, D.J. Thouless, J. Phys. C{\bf 5}, L124 (1973);
    J.Phys.C{\bf 6}, 1181 (1973); J.M. Kosterlitz, J. Phys. C{\bf 7}, 1046 (1974).

\bibitem{berezinskii}    V.L. Berezinskii, Sov. Phys. JETP {\bf 34}, 610 (1972)

\bibitem{nelsonkosterlitz}        D.R. Nelson, J.M. Kosterlitz, Phys.Rev.Lett. {\bf 39},
    1201 (1977)

\bibitem{caldeiraleggett}  A.O. Caldeira, A.J. Leggett, Ann.Phys. {\bf 149}, 374 (1983)

\bibitem{LiHo24}      C. Simon, D.M. Silevitch, P.C.E. Stamp, T.F. Rosenbaum, PNAS
                             {\bf 121}, 2315598121 (2024)





\bibitem{chan74}           M.H.W. Chan, J.D. Yanof, J.D. Reppy, Phys. Rev. Lett. {\bf 32},
    1347 (1974)

\bibitem{hallock21}        J.B. Hallock, J. Low Temp. Phys. {\bf 205}, 160 (2021)



\bibitem{kozikproksvis}    E.Kozik, N.V.Prokof'ev, B.Svistunov, Phys. Rev. B{\bf 73},
    092501 (2006)

\bibitem{thompsonstamp}    L.Thompson, P.C.E. Stamp, Phys. Rev. Lett. {\bf 108}, 184501
    (2012)

\bibitem{forsachwool}     S. Forstner, Y. Sachkou, M. Woolley, G.I. Harris, X. He, W.P.
    Bowen, C. Baker, New J.Phys. {\bf 21}, 053029 (2019)

\bibitem{harveyfetter}    K.C. Harvey, A.L. Fetter, J. Low  Temp. Phys. {\bf 11}, 473
    (1973)

\bibitem{vittocolemeinke}    E. Vittoratus, M.W. Cole, P.P.M. Meinke,  Can. J. Phys. {\bf
    51}, 2283 (1973)

\bibitem{rayfield}       We expect $\xi_0$ to be similar to the 3-d value $\xi_0 \sim 1.7 \times 10^{-10}$m (for which see Bowley et al. in ref. \cite{3d-expt} above, or G.W. Rayfield, F. Reif, Phys. Rev. {\bf 136}, 1194 (1964)). This is because its value is controlled by high-energy physics, largely unaffected by the dimensionality. 



\bibitem{faddeevpopov}       V.N. Popov, L.D. Faddeev, Sov. Phys. J.E.T.P. {\bf 20}, 890
    (1965); V.N. Popov, Theoretical and Mathematical Physics {\bf 11}, 565 (1972)


\bibitem{QPpair}          The argument showing that there can be no coupling between the
vortex and single quasiparticles is straightforward \citep{thompsonstamp}. The vortex is
a minimum action solution of the equations of motion, and so there can be no linear
coupling to fluctuations (ie., to quasiparticles) around it; the lowest-order coupling is quadratic.


\bibitem{arovas97}          D.P. Arovas, J.A. Freire, Phys. Rev. {\bf B}55, 1068 (1997)


\bibitem{frousteycoxstamp} J.Froustey, T.Cox, P.C.E. Stamp, to be published. The
     quasiparticle fields $\boldsymbol{A}\left(\boldsymbol{R}_{V},t\right)$ and $V_{V}
     \left(\boldsymbol{R}_{V},t\right) $, when coupled to the vortex,
    give rise to radiative corrections to the vortex dynamics: these can be evaluated
    both perturbatively and non-perturbatively (the latter using a functional eikonal
    expansion).



\bibitem{supplements}        See Supplementary Information

\bibitem{marchand}    D.Marchand, "Vortex Nucleation in a Superfluid", M.Sc. thesis (Univ.
    of British Columbia, 2006)

\bibitem{volovik}      G.E. Volovik, J.E.T.P. Lett {\bf 15}, 81, (1972)





\bibitem{suhl} H. Suhl, Phys. Rev. Lett. {\bf 14}, 226 (1965)

\bibitem{duan}    J.M. Duan, Phys. Rev. B{\bf 49}, 12381 (1994); see also J.M. Duan,
    Phys. Rev. B{\bf 48}, 333 (1993), and J.M. Duan, A.J. Leggett, Phys. Rev. Lett. {\bf
    68}, 1216 (1992)


\bibitem{popov}   V.N. Popov, Zh.E.T.F. {\bf 64}, 672 (1973), Sov.Phys. J.E.T.P. {\bf 37},
    341 (1973)

\bibitem{baym}      G.Baym, G.Chandler, J. Low Temp. Phys. {\bf 50}, 59 (1983)

\bibitem{niuaothouless} Q. Niu, P. Ao, D.J. Thouless, Phys. Rev. Lett {\bf 72}, 1706
    (1994); J.M. Duan, Phys. Rev. Lett. {\bf 75}, 974 (1995); and Q. Niu, P. Ao, D.J.
    Thouless, Phys. Rev. Lett. {\bf 75}, 975 (1995).


\bibitem{thouanglin}         D.J. Thouless, J.R. Anglin, Phys. Rev. Lett. {\bf 99}, 105301
    (2007)

















\bibitem{desrochersstamp} M.J. Desrochers, P.C.E. Stamp, to be published.




\bibitem{SQUID} Our analysis here is similar to that applied to SQUIDs; see J.Kurkijarvi,
    Phys. Rev. {\bf B6}, 832 (1972).



\bibitem{KeZhou}   One can also imagine fabricating a substrate with a region, far from any boundary, in which the van der Waals capillary interaction is different from its surroundings. In this region the vortex energetics would be different, and we could envisage doing experiments purely in this region, where all vortex nucleation would be intrinsic.


\bibitem{coleman}     C. Coleman, Phys. Rev. D{\bf 15}, 2929 (1977); S. Coleman, C.G.
    Callan, Phys. Rev. D{\bf 16}, 1762 (1977)

\bibitem{bailin80}    D. Bailin, A. Love, J. Phys. A{\bf 13}, L271 (1980). For a recent
    discussion of experiments on the superfluid He-3 A-B transition, see Y. Tian et al.,
    Nat. Comm. {\bf 14}, 148 (2023)

\bibitem{Qturb}    See, eg., E. Varga, V. Vadakkumbatt, A. J. Shook, P. H. Kim, J. P.
    Davis, Phys. Rev. Lett. {\bf 125}, 025301 (2020), and O.R. Stockdale et al., Phys.
    Rev. Research {\bf 2}, 033138 (2020).






\end{thebibliography}
\end{document}


\preprint{}
	
	\title{Vortex Tunneling in 2-Dimensional $^{4}$He Superfluid Films - Supplementary Information}

\maketitle

In this supplementary information we give further details of the derivations
of some of the results in the main body of the paper. We first give more details of the theoretical model for the 2-dimensional superfluid film and the vortices in it, in long wavelength approximation; and then we discuss the vortex energetics and tunneling rates for the various configurations discussed in the main text.

\vspace{10mm}


\section{The $^{4}$He Superfluid Film}


In both 2 and 3 dimensions, one can describe a low-temperature superfluid
$^{4}$He film by the bulk long-wavelength action {[}1{]}
\begin{equation}
S=\int dt\,\int d^{D}r\,\left[\frac{\hbar}{m_{0}}\rho\left(\boldsymbol{r},t\right)\dot{\Phi}\left(\boldsymbol{r},t\right)-\mathcal{H}\left(\boldsymbol{r},t\right)\right]\label{eq:Action-Bulk Longwavelength}
\end{equation}
where $D$ is the spatial dimension, $m_{0}$ the mass of a $^{4}$He
atom, $\rho\left(\boldsymbol{r},t\right)$ is the bulk mass density (which becomes a mass density per unit area in 2 dimensions, proportional to the thickness of the superfluid film), and $\Phi\left(\boldsymbol{r},t\right)$ the superfluid phase.

This long wavelength theory is valid for wavelengths $\gg \xi_o$, where $\xi_o$ is the ``healing length" in the system; for $^4$He films, $\xi_o \sim 1.7$ Angstroms (depending on the substrate), roughly half the interparticle spacing $a_o$ (which for a pressure of 25 bar is $a_o = 3.4$ Angstroms). The value for $\xi_o$ comes from experiments on critical velocities of negative ions in the 3-d superfluid \cite{rayfield,bowley}, but because $\xi_o$ is determined by high energy, short distance physics, we do not expect it to be changed by the spatial dimension. Indeed, our long-wavelength limit also assumes that we work at energy scales $\ll \Lambda_o = m_o c_o^2$. We can treat $\Lambda_o$ as an effective UV cutoff on the theory. We can estimate $\Lambda_o$ using $c_0\sim 240 \;m/s$. This gives a value of $\Lambda_o\sim 7\text{K}\equiv 150$ GHz. Thus our theory applies at length-scales $\gg 1\text{Å}$, for velocities $\ll 240\;m/s$, and frequency scales $\ll 150$ GHz.

The Hamiltonian density $\mathcal{H}\left(\boldsymbol{r},t\right)$ corresponding to this action is
\begin{equation}
\mathcal{H}\left(\boldsymbol{r},t\right)=\frac{1}{2}\frac{\rho\left(\boldsymbol{r},t\right)}{m_{0}^{2}}\left(\hbar\nabla\Phi\left(\boldsymbol{r},t\right)\right)^{2}-\epsilon\left(\boldsymbol{r},t\right)\label{eq:Hamiltonian Density}
\end{equation}
in which $\epsilon\left(\boldsymbol{r},t\right)=E\left(\boldsymbol{r},t\right)-E_{0}$
is the deviation of the energy density from that of the homogeneous
system (ie., it parameterizes energy fluctuations).

Let us write the mass density as \cite{thompsonstamp}
\begin{equation}
\rho\left(\boldsymbol{r},t\right)=\rho_{s}({\bf r}, t) + \eta\left(\boldsymbol{r},t\right)\label{eq:Mass Density}
\end{equation}
where $\rho_{s}({\bf r}, t)$ is the bulk $T=0$ superfluid density (so that $\rho_{s}({\bf r}, t) \rightarrow \rho_s$, a constant, for a film with uniform thickness). We can then expand the energy density functional as
\begin{equation}
\epsilon\left(\eta,\nabla\eta\right)=\frac{\hbar^{2}}{8m_{0}^{2}\rho}\left(\nabla\eta\right)^{2}+\frac{1}{2\chi\rho_{s}^{2}}\eta^{2}+\mathcal{O}\left(\eta^{4},\left(\nabla\eta\right)^{4}\right)\label{eq:Energy Density - Deviation}
\end{equation}
per unit area, where $\chi$ is the compressibility of the system. One can immediately derive the existence of a spectrum of low-energy phonons from this long-wavelength description; their velocity $c_o$ is given by
\begin{equation}
c^2_o \;=\; 1/\chi \rho_s \;\sim \; \Lambda_o/m_o
 \label{c-o}
\end{equation}

We note that for a superfluid $^{4}$He film, there are also important
``3rd sound'' film surface excitations {[}2,3{]}, and one must generalize the above energy functional to include surface waves. The superfluid
density and film thickness are controlled by a complicated interplay
between surface tension, van der Waals coupling to the substrate,
the ``bending energy'' which resists deformation of the superfluid
order parameter, and Bernouilli and centrifugal forces coming from
the superflow field itself {[}4,5{]}.

One can also have quasiparticle propagation along the $z$-axis for sufficiently thick films. For thin films these modes are blocked at low temperature. The lowest energy phonon propagating along $\hat{\bf z}$ has energy $\hbar c_0 (\pi/L_z)$; for a film of thickness $L_z = 50\xi_0$ this has energy $120$ mK, and for $L_z = 30 \xi_0$, it has energy $200$ mK.

More important, in films of finite thickness, quantized Kelvin modes, or Kelvons, can be generated along the length of what is now a vortex line extending from the substrate to the superfluid film surface. For a film of thickness $L_z$, these modes have frequency
\begin{equation}
\omega_K(k_z) \;=\; {\hbar k_z^2 \over 2 m_0} \ln \left( {1 \over \xi_o k_z} \right)
 \label{kelvon}
\end{equation}
where the allowed wave-vectors along the line are $k_z = (2n+1)\pi/L_z$. For a film thickness $L_z = 10\xi_0$, one then finds that the lowest ``zero point" Kelvon mode has frequency $\sim 20.8$ GHz, equivalent to a temperature $\sim 1$K; for a film thickness $L_z = 30 \xi_0$, this energy falls to $\sim 232$ mK, and for $L_z = 50 \xi_0$, it becomes $\sim 104$ mK. Thus Kelvon modes can be ignored for sufficiently low $T$.

At first glance these results indicate that we can ignore entirely the effect of quasiparticle excitations if we work at sufficiently low temperatures. However one has to be careful here. Quasiparticles have several effects on vortex dynamics, viz.,

(i) The ``radiative corrections", caused by emission and absorption of both real and virtual quasiparticles, will renormalize all vortex properties \cite{frousteycoxstamp}, even at $T=0$ (including the hydrodyanmic effective mass of a vortex); real quasiparticles at finite $T$ will cause dissipative friction in the vortex motion.

(ii) Even at $T=0$, purely quantum processes like vortex tunneling rates will be renormalized by the vortex/quasiparticle coupling; above a crossover temperature, the tunneling will switch to thermal activition, whose temperature dependence also depends on the vortex/quasiparticle coupling.

Once we have discussed the details of this coupling (see below) we will see that we expect its effects to be small, both on mass renormalization, and, in the low temperature range specified above (below $\sim 0.6$K), on vortex tunneling. Thus in the present paper we will not be including the effect of quasiparticles on the vortex dynamics.

We conclude that, because the effects of the vortex/quasiparticle coupling are small, and for sufficiently thin films and for temperatures less than $\sim 0.6$K, we can ignore 3-dimensional excitations in the superfluid film, and treat it as genuinely 2-dimensional. However, we still need to include vortices in the picture.

\vspace{10mm}


\section{ Vortices in Infinite Superfluid Films}


It is important to distinguish two cases, viz., (i) the case where the 2-d superfluid is unbounded, so that we can ignore the interaction of the vortices with any boundary; and (ii) the case where the vortices do interact with the boundary, and with any image vortices created by the boundary. In this section II we deal with the first case

\subsection{Single Vortices in Infinite Superfluid Films}

A vortex in a superfluid is an excitation quite different from the quasiparticle - it is a {\it quantum soliton}, and must be handled accordingly. For thin films, in our long wavelength description, we can treat
the vortex core (which has radius $\sim \xi_o$), as a point at position $\boldsymbol{R}_{V}\left(t\right)$.

However, the vortex flow field extends over long ranges. To describe this flow field, we follow Thompson and Stamp {[}6{]}, and write
$\rho\left(\boldsymbol{r},t\right)$ and $\Phi\left(\boldsymbol{r},t\right)$
around a vortex having a directed quantized circulation $\boldsymbol{\kappa}=\pm\hat{\boldsymbol{z}}(h/m_{0})$. We then write
\begin{equation}
\begin{cases}
\rho\left(\boldsymbol{r},t\right) & = \;\; \rho_{s}^{V}\left(\boldsymbol{r}-\boldsymbol{R}_{V}\left(t\right)\right)+\eta\left(\boldsymbol{r}-\boldsymbol{R}_{V}\left(t\right)\right)\\
\Phi\left(\boldsymbol{r},t\right) & = \;\; \Phi_{V}\left(\boldsymbol{r}-\boldsymbol{R}_{V}\left(t\right)\right)+\phi\left(\boldsymbol{r}-\boldsymbol{R}_{V}\left(t\right)\right)
\end{cases}\label{eq: 2d Film Long Wavelength Density and phase.}
\end{equation}
Here the subscript/superscript ``V" labels the vortex contribution, and the deviations $\eta,\phi$ come from quasiparticles.

In an infinite 2-d superfluid system, with no boundaries, the action
for a vortex with hydrodynamic mass $M^o_{V}$, in interaction with
the quasiparticles, and in the absence of any coupling to a boundary, is then {[}6,7{]}
\begin{equation}
S_{V}=\int dt\,\left[\frac{1}{2}M^o_{V}\left(\dot{\boldsymbol{R}}_{V}\left(t\right)-\boldsymbol{v}_{s}^{0}\right)^{2}+\boldsymbol{A}_{v}\left(\boldsymbol{R}_{V}\right)\cdot\left(\dot{\boldsymbol{R}}_{V}-\boldsymbol{v}_{s}^{0}\right)-V_{V}\left(\boldsymbol{R}_{V}\right)\right]
 \label{eq:Action - Vortex with hydrodynamic Mass}
\end{equation}
where $\boldsymbol{v}_{s}^{0}$ is the background superflow velocity
(ie., that part of $\boldsymbol{v}_{s}({\bf r})$ which is independent of the vortex).

In this action, the gauge potential
$\boldsymbol{A}_{V}(\boldsymbol{R}_{V})$ is given by the sum
\begin{equation}
\boldsymbol{A}_{V}(\boldsymbol{R}_{V})  =\boldsymbol{A}_{0}(\boldsymbol{R}_{V})+\boldsymbol{A}_{QP}\left(t\right)
 \label{eq:AV}
\end{equation}
where $\boldsymbol{A}_{0}\left(\boldsymbol{R}_{V}\right)$ is the Magnus potential
\begin{equation}
\boldsymbol{A}_{0}\left(\boldsymbol{R}_{V}\right)  =\frac{1}{2}\rho_{s}\left(\boldsymbol{R}_{V}\times\boldsymbol{\kappa}\right)
 \label{eq:A0}
\end{equation}
and $\boldsymbol{A}_{QP}\left(t\right)$ is that part of $\boldsymbol{A}_{V}\left(\boldsymbol{R}_{V}\right)$ arising from quasiparticle pair excitations in the normal fluid:
\begin{eqnarray}
\boldsymbol{A}_{QP}\left(t\right)=\frac{1}{2}\boldsymbol{J}_{QP}\left(t\right) & =&\frac{1}{2}\int d^{2}r\,\boldsymbol{j}_{qp}\left(\boldsymbol{r},t\right)\\
 & \equiv & \frac{1}{2}\int d^{2}r\,\frac{\hbar}{m_{0}}\left[\eta\nabla\phi-\phi\nabla\eta\right]
 \label{eq:AQP}
\end{eqnarray}
and is a description of a quasiparticle pair current, with pair current density $\boldsymbol{j}_{qp}\left(\boldsymbol{r},t\right)$ (in the main text we suppress the ``QP" and ``qp" subscripts).

The other quasiparticle term in the action is the potential term
\begin{equation}
V_{V}\left(\boldsymbol{R}_{V}\right)=\frac{\hbar^{2}}{2m_{0}^{2}}\int d^{2}r\,\nabla\Phi_{V}\left(\boldsymbol{r},t\right)\cdot\left(\eta\nabla\phi-\phi\nabla\eta\right)\label{eq:Vv}
\end{equation}
which describes the interaction between the quasiparticles and the long-range flow field of the vortex.

We see that the Magnus term $\boldsymbol{A}_{0}\left(\boldsymbol{R}_{V}\right)$
exists in the absence of quasiparticles, and is always proportional to $\rho_S$. All
the other terms, involving gradients of the density and phase fluctuations $\eta,\phi$ respectively, explicitly involve quasiparticles (phonons, 3rd sound, etc). The terms
$\boldsymbol{A}_{QP}\left(t\right)$ and $V_{V}\left(\boldsymbol{R}_{V}\right)$ explicitly describe vortex-quasiparticle
interactions.

The  long-wavelength Hamiltonian corresponding
to this action is just
\begin{equation}
\mathcal{H}_{V}\left(\boldsymbol{R}_{V},\boldsymbol{P}_{V}\right)=\frac{1}{2M^o_{V}}\left(\boldsymbol{P}_{V}-\boldsymbol{A}_{V}\left(\boldsymbol{R}_{V}\right)\right)^{2}+V_{V}\left(\boldsymbol{R}_{V}\right)\label{eq:Hamilgonian Density - Effective long-wavelength}
\end{equation}
where $\boldsymbol{P}_{V}=-i\hbar\nabla_{\boldsymbol{R}_{V}}$ is
the vortex momentum operator (and $\boldsymbol{P}_{V}-\boldsymbol{A}_{V}\left(\boldsymbol{R}_{V}\right)$ is the gauge invariant vortex momentum).

\vspace{2mm}

We see that if we completely ignore the quasiparticles, we are left with
a ``bare vortex'' action given by
\begin{equation}
S_{V}^{0}=\int dt\,\frac{1}{2}\left[M^o_{V}\left(\dot{\boldsymbol{R}}_{V}-\boldsymbol{v}_{s}^{0}\right)^{2}+\rho_{s}\left(\boldsymbol{\kappa}\times\boldsymbol{R}_{V}\right)\cdot\left(\dot{\boldsymbol{R}}_{V}-\boldsymbol{v}_{s}^{0}\right)\right]\label{eq:Action - Bare Vortex}
\end{equation}
However in all problems we are interested in, we will deal either with an interacting vortex/anti-vortex pair, in an otherwise unbounded system; or we will deal with a bonded system in which a vortex interacts with image charges. Thus the result just given for an isolated vortex is not immediately useful - one always deals with multiple interacting vortices.

\subsection{Vortex/Anti-Vortex Pair in Infinite Superfluid Films}

The calculations above are easily generalized to the case of a multi-vortex system; however the physics is now very different, because each vortex exerts forces on the others, and new length scales are introduced into the system.

We will focus here on the vortex/anti-vortex pair system. As discussed in the main text, we assume
a uniform background superflow field with velocity
$\boldsymbol{v}_{S}$$\left(\boldsymbol{r}\right)\rightarrow\boldsymbol{v}_{0}$, and we wish to write down an effective action for the pair of vortices in this flow field.

We therefore define vortex/anti-vortex coordinates $\left\{ \boldsymbol{R}_{j}\right\} $,
with $j\in\left\{ 1,2\right\} $ so that the effective action given in the main text for a single vortex now becomes, for the vortex/anti-vortex pair:
\begin{equation}
S_{\text{eff}}^{V}\left[\left\{ \boldsymbol{R}_{j},\dot{\boldsymbol{R}}_{j}\right\}
\right]=\int dt\sum_{j\in\left\{ 1,2\right\}
}\left[\frac{1}{2}M_{V}\left(\boldsymbol{R}_{j}\right)\dot{R}_{j}^{2}+
\dot{\boldsymbol{R}}_{j}\cdot\boldsymbol{A}_{V}\left(\boldsymbol{R}_{j},t\right)-
V_{V}\left(\boldsymbol{R}_{V},t\right)\right]
 \label{eq:Vortex/anti-vortex Effective Action}
\end{equation}

In line with what was done before, we will drop the quasiparticle terms, ie., we will assume that $\boldsymbol{A}_{V}\left(\boldsymbol{R}_{j},t\right) \rightarrow  \boldsymbol{A}_{0}\left(\boldsymbol{R}_{j}\right)$, and $V_{V}\left(\boldsymbol{R}_{V},t\right) \rightarrow 0$. We now define sum and difference coordinates
$\boldsymbol{Q}=\frac{1}{2}\left(\boldsymbol{R}_{1}+\boldsymbol{R}_{2}\right)$
and $\boldsymbol{r}_{12}=\left(\boldsymbol{R}_{1}-\boldsymbol{R}_{2}\right)$. In a fairly lengthy derivation, one can then rewrite the system pair system action in terms of these new coordinates \cite{laraThesis}, in the form
\begin{equation}
S_{\text{eff}}^{V} (\boldsymbol{Q}, \boldsymbol{r}_{12}; \dot{\boldsymbol{Q}},
\dot{\boldsymbol{r}}_{12} ) \;=\; \int dt \left[\frac{1}{2} M_{\alpha \beta} \, \dot{Q}
^{\alpha} \dot{r}_{12}^{\beta}
\;-\;  {\cal T} (\boldsymbol{Q}, \boldsymbol{r}_{12}) \right]
 \label{eq:Vortex/anti-vortex S}
\end{equation}
where $M_{\alpha \beta}$ is an effective mass tensor for the vortex/anti-vortex pair system, and where ${\cal T} (\boldsymbol{r}_{12})$ is an effective potential coupling the vortex/anti-vortex pair.

The dynamics in the summed coordinate $\boldsymbol{Q}=\frac{1}{2}\left(\boldsymbol{R}_{1}+\boldsymbol{R}_{2}\right)$ is of no interest to us - we want the effective potential in the vortex/anti-vortex separation coordinate $\boldsymbol{r}_{12}=\left(\boldsymbol{R}_{1}-\boldsymbol{R}_{2}\right)$. This allows us to consider the scenario discussed in the main text, in which a vortex/anti-vortex pair
nucleates far from any boundary, in the presence of a background flow
$\boldsymbol{v}_{s}^{0}$. The superflow energy can be found either by direct construction, or by noting that the problem is completely equivalent to the nucleation
of a vortex at the surface of a flat wall, with an anti-vortex being
provided by the image potential inside the wall. Thus, the flow is that
of a dipole superposed on the background superflow. If the coordinates
of the vortex/anti-vortex pair are $\boldsymbol{R}_{1}$ and $\boldsymbol{R}_{2}$,
then the superflow energy is just a simple combination of the Magnus interaction with the external flow field and the logarithmic interaction between the vortex/anti-vortex pair. One gets the known result \cite{donnelly}:
\begin{equation}
\mathcal{T} ({\bf r}_{12} )=\frac{\rho_{s}\kappa^{2}}{4\pi} \ln\left|\frac{r_{12}}{\xi_{0}}\right|-\frac{1}{2}\rho_{s}\kappa\boldsymbol{v}_{s}^{0}\cdot\boldsymbol{r}_{12}
 \label{eq:Energy - Vortex/Antivortex Pair}
\end{equation}
where $\boldsymbol{r}_{12} = \left(\boldsymbol{R}_{1}-\boldsymbol{R}_{2}\right)$.
The linear term which pulls the vortex apart from the anti-vortex comes directly from the Magnus force. It is maximized when $\boldsymbol{r}_{12}$ is perpendicular to $\boldsymbol{v}_{s}^{0}$, and this direction then provides the MPEP (Maximum Probability Escape Path) for vacuum tunneling of the vortex/anti-vortex pair.

\vspace{2mm}

$\qquad\qquad\qquad\qquad\qquad\qquad\qquad\qquad\qquad\qquad$------------------------------

\vspace{2mm}

It is useful to compare this formulation of the vortex problem with others given in the past. Almost all previous discussions of the vortex/quasiparticle system have been phenomenological, and have treated the vortex as a classical object - such descriptions go back to Hall and Vinen, and Iordanski \citep{hallV,iordanskii}. Obviously such treatments cannot deal with any quantum dynamics of vortices, including vortex tunneling. Notable exceptions include:

(i) Tunneling calculations of Volovik \cite{volovik}, Sonin \cite{sonin}, and Donnelly \cite{donnelly}, in which extrinsic tunneling in various simplified 3-dimensional geometries - having a high degree of symmetry - are considered. In these calculations a hydrodynamic model for the superfluid is considered, and the vortex mass is taken to be either zero or a small constant describing the vortex core mass. Quasiparticles do not enter the picture.

(ii) Phenomenological tunneling calculations for simplified 3-dimensional geometries by Varoquax et al. \cite{varoquaux}, in which the mass of the vortex is taken to be the $^4$He atomic mass $m_0$, and quasiparticle dissipation is introduced by hand, assuming an Ohmic Caldeira-Leggett model.

(iii) A mapping by Arovas et al. \cite{arovas} of the 2-d vortex/quasiparticle system to a $2+1$-d QED, in which the photons represented single phonon quasiparticles; and the use of this model, with non-Ohmic dissipation, to calculate vortex tunneling with dissipative corrections to get a frequency-dependent vortex mass. The vortices were given a constant bare core mass.

(iv) An attempt by Thouless et al. \cite{thouless} to derive a quantum equation of motion for a vortex coupled to quasiparticles, in which particular attention was paid to the Magnus and Iordanskii forces.

(v) Derivation by Thompson and Stamp \cite{thompsonstamp} of a fully quantum equation of motion for a vortex in a 2-dimensional superfluid, in which the vortex was shown to only interact with quasiparticle pairs, representable as a gauge field, and where both the hydrodynamic mass of the vortex and quantum dissipative contributions to the dynamics were derived. Extension of this work to multi-vortex systems, and to include radiative corrections (for which see refs. \cite{laraThesis,frousteycoxstamp}).

\vspace{2mm}

We see that these different approaches all differ in key aspects from each other. The key differences between the framework used here and those employed in (i)-(iv) above are:

1.	We have a position-dependent mass, and all our tunneling calculations are done with this mass. Radiative corrections to the mass are small.

2.	We consider both intrinsic and extrinsic nucleation – intrinsic tunneling processes have never been considered before.

3.	Our microscopic action is different from these authors (principally because our gauge field describes quasiparticle pairs, and because we have a position-dependent vortex mass).

\vspace{10mm}


\section{Vortex Energetics in a Film with Boundaries}


As soon as we introduce a boundary, any vortex will interact with image vortices created by the boundary - this introduces new length scales into the problem.
In our long-wavelength treatment, the vortex core can be treated as
a point, and the flow field $\boldsymbol{v}_{s}\left(\boldsymbol{r}\right)$
is then simply a 2-dimensional field on some 2-d domain. We can then
use complex variable theory to model the flow.

We write $\boldsymbol{v}_{s}=\nabla\varphi\left(z\right)$, where $\varphi\left(z\right)$ is a phase function,
where $z=x+iy$ is the coordinate on the plane, and $\nabla^{2}\varphi=0$
everywhere except at points where we have vortices. We can also write
a stream function $\psi\left(z\right)$, such that one has $\boldsymbol{v}_{s}=\nabla\times\psi\left(z\right)$,
and thereby define a complex velocity potential $\Omega\left(z\right)=\varphi\left(z\right)+i\psi\left(z\right)$;
we then have $\partial_{z}\Omega\left(z\right)=v_{s}^{x}-iv_{s}^{y}$.

The total flow energy of the system is then just given by
\begin{equation}
{\cal T} \;=\; \tfrac{1}{2}\rho_s \int d^2 r \; v_s^2 ({\bf r})
 \label{calT-z}
\end{equation}
where the integration extends over the entire system, bounded or otherwise, with some prescription being given for the contribution from the vortex cores (which here are just point sources). The great advantage of this formulation is that it automatically includes all cotributions to the energy, including those deriving from Magnus forces as well as the interaction with image vortices.

For a single vortex at the origin, in the absence of any boundary, one simply has for the phase and stream functions the results
\begin{equation}
\begin{cases}
\varphi\left(z\right) & = (\kappa/2\pi)\tan^{-1}\left(y/x\right)= (\kappa/2\pi)\Theta\\
\psi\left(z\right) & = - (\kappa/2\pi) \ln \,r
\end{cases}\label{eq: Complex Velocity Potential}
\end{equation}
where $r^{2}=x^{2}+y^{2}$, $\kappa= h/m_{0}$ is the quantum
of circulation, and $\Theta$ the polar angle; the velocity potential
is then
\begin{equation}
\Omega\left(z\right)=-i\frac{\kappa}{2\pi}\ln \,z
 \label{eq:Velocity Potential}
\end{equation}
For an infinite system the energy of this isolated vortex is also infinite; as is well known, it diverges logarithmically with the size $R_o$ of the system, according to
\begin{equation}
{\rho_s \kappa^2 \over 4\pi} \; \ln |R_o/\xi_o|
 \label{E-1vortex}
\end{equation}

Let us now treat the two cases of interest for the main text.

\subsection{ Vortex in a \textquotedblleft Cylinder\textquotedblright}

We assume the geometry shown in the main text, with a vortex of circulation $\kappa$, situated at a distance $R_v$ along the $\hat{x}$ axis from the centre of a circular container of radius
$R_{0}$. We then have a velocity potential at position $z=x+iy$ given by
\begin{equation}
\Omega\left(z\right)=-i\frac{\kappa}{2\pi}\ln\left(\frac{z-r}{z-R_{0}^{2}/R_V}\right)\label{eq:Velocity Potential - Cylinder}
\end{equation}
and, integrating the superflow kinetic energy $\frac{1}{2}\rho_{s}v^2_{s}\left(z\right)$
over the entire system, with a short distance cutoff $\xi_{0}$ at the vortex
core, we get the total energy associated with the interaction between the vortex and its image anti-vortex to be
\begin{equation}
\mathcal{T}_{0}\left(r\right)=\frac{\rho_{s}}{4\pi}\kappa^{2}\ln\left(\frac{R_{0}^{2}-R_V^{2}}{R_{0}\xi_{0}}\right)
 \label{eq:Energy-Cylinder}
\end{equation}

The above result assumes that the only flow inside the ciruclar domain
is caused by the vortex. However, we can also consider a rotating
circular container, rotating with angular velocity $\boldsymbol{\omega}$
(so that the relative velocity of the background superflow at the circular boundary to the boundary wall,
in the absence of any vortex, is $\boldsymbol{v}_{s}^{0}\left(R_{0}\right)=\omega R_{0}$).

In this case, the superflow energy is given, in the frame of the wall,
by
\begin{align}
\mathcal{T}_{\omega}\left(r\right) & =\mathcal{T}_{0}\left(r\right)-\omega L\nonumber \\
 & =\frac{\rho_{s}}{4\pi}\kappa^{2}\ln\left(\frac{R_{0}^{2}-R_V^{2}}{R_{0}\xi_{0}}\right)-\kappa\rho_{s}\omega\left(R_{0}^{2}-R_V^{2}\right)
 \label{eq:Energy - Cylinder with Rotation}
\end{align}
where $L$ is the angular momentum of the superfluid in the rotating
frame. These results are also well known \cite{donnelly}.

\subsection{ Semi-Circular Projection on the Flat Wall}

Now consider the geometry discussed at length in the main text, viz., superflow past a semi-circular projection of radius
$r_{0}$, attached to a flat wall. We assume a superflow velocity
$\boldsymbol{v}_{s}^o$, parallel to the wall, at distances $\gg r_{0}$
from the projection (See figure in the main text).

Let us place the origin at the
centre of the semi-circular projection. We then have, for a point at $z$ in the complex plane, and with a
vortex at position $z_{0}$ in that plane, the velocity potential
\begin{equation}
\Omega\left(z\right)\;\;=\;\; v_{s}^o\frac{r_{0}^{2}}{z}\;-\; i\frac{\kappa}{2\pi}\left[\ln\left(\frac{z-Z_{0}}{z-\overline{Z}_{0}}\right)-\ln\left(\frac{z-z_{0}}{z-\overline{z}_{0}}\right)\right]
 \label{eq:Velocity Potential - Bump}
\end{equation}
where we have written $Z_{0}=R_{V}e^{i\alpha}$ as the complex coordinate of the vortex
with $R_{v}\geq r_{0}$, at angle $\alpha$ from the wall. The angle $\alpha$ is then that between the superflow velocity $\boldsymbol{v}_{s}^{0}$ at infinity, ie.,
along $\hat{\boldsymbol{x}}$, and the radius vector to the point
$z$. Then $z_{0}$ is the complex coordinate of
the image vortex inside the semi-circular projection, with $z_{0}=\left(r_{0}^{2}/R_{V}\right)\exp\left\{ i\alpha\right\} $.
Finally, $\overline{Z}_{0}$ and $\overline{z}_{0}$ are the conjugates
of $Z_{0}$ and $z_{0}$ respectively.

The superflow energy for this system can then be found by very lengthy
integrations. We have done this using two different methods, viz.,

(i) by straightforward integration of the superfluid energy, ie., by evaluation of $\tfrac{1}{2} \rho_s \int d^2r \; v^2_s({\bf r})$ over the entire domain. The algebra in this method turns out to be extremely unwieldy.

(ii) we transform this integral into a set of 4 different integrals, involving the phase and stream functions and their derivatives, over the 4 different bounding sections of the domain. This also leads to very lengthy expressions \cite{furtherC}

Doing the calculation by both methods allows us to check the answer. This final result is a total energy describing the interaction between the vortex and the boundary of the system, given by
\begin{eqnarray}
\mathcal{T}( {\bf R}_V)\;\;&=&\;\;\frac{\rho_{s}\kappa^{2}}{8\pi}\left\{
\ln\left(\frac{4R_{V}^{2}\sin^{2}\alpha}{\xi_{0}^{2}}\right)
- \ln\left[\left(\frac{R_{V}^{2}+r_{0}^{2}}{R_{V}^{2}-r_{0}^{2}}\right)^{2}\sin^{2}\alpha+\cos^{2}\alpha\right]  \right\}  \nonumber\\
&& \qquad\qquad\qquad - {\rho_s\kappa v_s^o \over 4} \left\{ \left(\frac{R_{V}^{2}- r_{0}^{2}}{R_{V}}\right)\sin\alpha \,+\,  {8 r_0 \over \pi} \tan^{-1} \left(\frac{r_{0}R_{V}\sin \alpha}
 {R_V^{2}-r_{\text{0}}^{2}}\right) \right\}  \;-\; {\pi \rho_s r_0^2 \over 4} (v_s^o)^2
  \label{eq:Energy-Bump}
\end{eqnarray}
which is composed of the following terms:

(i) the first two terms, proportional to $\kappa^2$, which describe interactions between the vortex and its two mirror anti-vortices; the first term is the $z_o - \overline{z}_{0}$ interaction and the second is the $Z_o - \overline{Z}_{0}$ interaction;

(ii) the next two terms, proportional to $\kappa$, describe the Magnus interaction between the external supercurrent and the vortex; the first of these is what one gets for a uniform supercurrent, and the second describes the angular corrections to this coming from the distortion of the superflow around the protuberance;

(iii) the last term, independent of $\kappa$, which is of no real interest here. In the frame of the wall we would actually get the result $(\pi \rho_s r_0^2 (v_s^o)^2/ 4) [R_0^2 - r_o^2]$ for this term, which would be the bulk energy of the uniform superflow in a region of area $A = \pi [R_0^2 - r_o^2]/2$, ie., a half-circle of radius $R_0$ with a semicircle of radius  $r_0$ excised. However the expression here is calculated in the frame of the uniformly moving superfluid.

It is interesting to note certain key features of eqtn. (\ref{eq:Energy-Bump}). First,
we observe that for large $R_V \gg r_o$, the angular corrections to the second ``Magnus"
term (proportional to $\kappa$) die off, and we are left with the uniform supercurrent
interaction. Second, the interaction term $\sim \kappa^2$ is almost everywhere
completely dominated by the term $\sim \ln |R_v \sin \alpha/\xi_0|$, and we can
usually ignore the other small corrections to this.

Finally, the energy $\mathcal{T}({\bf R}_V)$ acts like a potential - plotted in the
main text - through which tunneling nucleation can take place. It is clear from the
figure in the main text, as well as from detailed analysis of (\ref{eq:Energy-Bump}),
that the tunneling action is minimized by following the path along the line $
\alpha=\pi/2$. Again, this path is the path along which the (negative) effect of
the Magnus term is maximized. This path is then the MPEP (Maximum Probability Escape
Path).

\vspace{10mm}


\section{Tunneling Rates}


Here we describe how the results for the tunneling rates are obtained. The form for the tunneling action is not standard, but we show that it can be handled using conventional WKB/tunneling methods.

\subsection{Effective Mass and WKB/Instanton Tunneling Rates}

In ordinary WKB theory, the tunneling rate for a particle out of some
potential well is given by \cite{LL}
\begin{equation}
\Gamma_{\text{WKB}}=\Omega_{0}\exp\left\{ -\frac{1}{\hbar}\int_{a}^{b}dx\,\left[2m\left|V\left(x\right)-E\right|\right]^{\frac{1}{2}} \right\}
  \label{eq: WKB Tunneling Rate}
\end{equation}
where $m$ is the mass of the particle, $E$ its energy, and $V\left(x\right)$
is the potential through which it tunnels; the prefactor $\Omega_{0}$ is
an attempt frequency, and $a$ and $b$ are end-points. We note that for zero-temperature tunneling out of a potential well, one picks $E = \tfrac{1}{2} \hbar \omega_0$, where $\omega$ is the small oscillation frequency in the well, and this determines the turning points (entry and exit points) $a$ and $b$. One then has $\Omega_o \sim \omega_o$.

However in dealing with the tunneling of the vortices within the superfluid,
we are faced with two complications:

\begin{enumerate}
\item We are using a long-wavelength description of the tunneling, valid
only for lengthscales $>\xi_{0}$. In weakly-coupled Bose superfluids one can use a Gross-Pitaevskii
description of the short-range physics {[}6,8,9{]}, but for the strongly-coupled $^{4}$He superfluid this is quite inadequate. We then do not have a reliable
description of the superfluid at length scales $\sim\xi_{0}$, or for
timescales $\sim \hbar/m_{0}c_{0}^{2}$. This makes it hard to determine $E$ and $\Omega_o$, and hence $a$ and $b$. Luckily the tunneling
energetics is dominated by superflow over lengthscales $\gg\xi_{0}$,
as can be seen in equations $\text{\eqref{eq:Energy-Cylinder}},
 \text{\eqref{eq:Energy - Cylinder with Rotation}, \eqref{eq:Energy-Bump}, and \eqref{eq:Energy - Vortex/Antivortex Pair}}$.
Thus we can treat the effect of the short-range physics in terms of
a undetermined constant in the tunneling rate; this is in fact the
prefactor $\Omega_{0}$, as we will see below. There is a similar uncertainty in the values of the entry and exit points $a$ and $b$.

In fact, $\Omega_0$ is given roughly by the small oscillation frequency of a $^{4}$He atom in the vortex core, and the entry point $a$ is then $a \sim \xi_o$. This means that $\Omega_0$ is somewhat less than $\Lambda_0$, the high-energy cutoff for our theory (we recall that $\Lambda_0 \sim 150$GHz, whereas $\Omega_o \sim 50$GHz).  It also means that $a$ is smaller then the lengthscales for which the theory is valid, and smaller even than the distance scale set by the interparticle spacing (which is roughly 4 times larger than $\xi_o$). We see that it is just as sensible to fix $a=0$ in the theory as it is to fix $a = \xi_o$, provided that we adjust $\Omega_o$ accordingly.

\vspace{2mm}

\item The effective mass of the vortex is not a simple quantity, nor is
it a constant - in fact it increases as the tunneling process proceeds.
The topic of vortex effective masses has been very controversial over
the years {[}10{]}, and one can distinguish several different effective
masses, depending on the context. A full discussion of these different masses, particularly when more than one vortex is involved, is extremely lengthy \cite{frousteycoxstamp,laraThesis}. If we have an isolated vortex, things simplify \citep{effectivemass}; but in our case, we either have  a boundary in the system, or multiple vortices. The effective mass itself becomes a dynamic quantity, and varies with the position of the vortex.

\end{enumerate}

In what follows, we give discuss both the effective mass and the WKB tunneling rates, both for the case of a vortex near a boundary, and for the vortex/anti-vortex pair far from any boundary. We also briefly discuss the topic of radiative corrections to the effective mass.

\subsubsection{Dynamic Effective mass}

A result which was demonstrated in various ways by Suhl, Popov, Duan and Thouless
and Anglin \citep{effectivemass}, is that the dynamic effective mass of an {\it isolated
vortex} can be written as
\begin{equation}
M_{V} \;=\; {\mathcal{T} \over c_{0}^{2}} \;\;\;\; \rightarrow \;\;\;\;  {\rho_s \kappa^2 \over 4\pi c_o^2} \; \ln |R_o/\xi_o|  \qquad\qquad\qquad (isolated \; vortex)
 \label{eq:Dynamic Effective Mass}
\end{equation}
where ${\cal T}$ is the vortex energy, and $R_o$ is some upper distance cutoff (for an unbounded system, $R_o \rightarrow \infty$). This result for the hydrodynamic mass ignores short distance contributions
to the mass, of the kind discussed by Baym and Chandler \citep{effectivemass}; but
these are small corrections to the logarithmic factor.

However, if we have a boundary in the system, or if we have more than one vortex, the result (\ref{eq:Dynamic Effective Mass}) is no longer true. We stress that

(i) the analyses given by Duan and others \citep{effectivemass}, in which the main contribution to the vortex mass comes from the non-local change in areal mass density $\rho({\bf r})$ caused by the vortex motion, are still correct.

(ii) However in a multi-vortex system (ie., one in which there is more than one vortex, or a vortex/anti-vortex pair, or a vortex interacting with image vortices) then the effective mass becomes rather complex. One needs to use an effective mass tensor to describe the vortex motion, and this tensor depends on the position of each of the vortices; a general analysis is very involved \citep{frousteycoxstamp,laraThesis,MJdR+PCES}.

(iii) In a situation where one has high symmetry, the algebra simplifies greatly, and one term completely dominates. In the two cases we will study, this turns out to be the case.

With all this in mind, we turn to the two cases of interest here, where the results do indeed simplify because the relevant tunneling paths are simple. We have \cite{furtherC}

\vspace{3mm}

(a) {\it Semicircular Protuberance Problem}: In this case the WKB tunneling path is along the straight line $\alpha = \pi/2$. Thus we have a 1-dimensional tunneling problem, and we find a very simple result for the effective mass tensor - it becomes diagonal in the basis where one defines sum and difference coordinates for the vortex and its anti-vortex images (compare discussion after eqtn. (\ref{eq:Vortex/anti-vortex S})). We are only interested in the effective mass as a function of the coordinate $ R_V-r_o$, and one then finds that the effective mass for a displacement along the tunneling path takes the simple form
\begin{eqnarray}
M_V(R_V) \;\;&=&\;\; {\rho_s \kappa^2 \over 4\pi c_o^2} \,\left\{  \ln \left| {R_v - r_0 \over \xi_o} \right|  \;\;\;+\;\;\;    O \left( \ln \left|{R_V - r_0 \over r_0}\right|, \ln\left|{R_V \over r_0} \right| \; + \; etc.\right) \right\} \nonumber\\
&\sim &\;\; {\rho_s \kappa^2 \over 4\pi c_o^2} \,  \ln \left| {R_v - r_0 \over \xi_o} \right|
 \label{Meff-bump}
\end{eqnarray}
with the leading term becoming exact when $(R_v - r_0) \ll r_0$ (which is the case of interest for tunneling off the protuberance). This is a long wavelength result, not valid if $(R_v - r_0) \sim \xi_0$.

\vspace{3mm}

(b) {\it Vortex/Anti-Vortex Tunneling}: In this case the tunneling path is also very simple - the vortex and anti-vortex separate along the path perpendicular to the external current, along the direction of the Magnus force. Again we have a 1-dimensional tunneling problem, and again the effective mass tensor is diagonal in the basis of sum and difference coordinates for the vortex and anti-vortex. The effective mass for the relative coordinate is easily obtained (the argument exactly parallels that of Duan \citep{effectivemass}), and is just
\begin{equation}
M_V(R_V) \;\;=\;\; {\rho_s \kappa^2 \over 4\pi c_o^2} \, \ln \left| {r_{12} \over \xi_o} \right|
 \label{Mv-V-aV}
\end{equation}
where, as with all the other expressions derived here, this result is a long-wavelength result (ie., it cannot be taken seriously if $r_{12} \sim \xi_o$).

\vspace{2mm}

$\qquad\qquad\qquad\qquad\qquad\qquad\qquad\qquad\qquad\qquad$------------------------------

\vspace{2mm}

Finally, let us briefly consider radiative corrections to these results. This is a very technical topic, and is dealt with in detail elsewhere (see ref. \citep{frousteycoxstamp}, and also refs. \citep{thompsonstamp,laraThesis}). As noted already in the main text, a key feature of the vortex-quasiparticle interaction is that it must, at lowest order, be quadratic in the quasiparticle variables. This is because the vortex is a minimum action soliton solution to the equations of motion, and so all fluctuations around it have to be quadratic in the fluctuations - these fluctuations are just the quasiparticle variables themselves \citep{thompsonstamp}.

To see this formally one consider a vortex in uniform motion, in some direction $\hat {\bm n}_0$ with respect to the superfluid \cite{thompsonstamp}. This is equivalent to a ``zero mode" deviation from the static superfluid profile of form
\begin{equation}
 \label{zeromode}
\phi_0 = \nabla \Phi_V \cdot \hat {\bm n}_0      \qquad\qquad\qquad
\eta_0 = \nabla \rho_s^V \cdot \hat {\bm n}_0
\end{equation}
where $\Phi_V$ and $\rho_s^V$ were defined in eqtn. (\ref{eq: 2d Film Long Wavelength Density and phase.}).

Expanding in the vortex velocity $ {\bm \dot
{\bm R}}_V$ (assumed small compared to the sound velocity $c_0$), one then finds new contributions to the action of form
$\Delta S_{int} = \Delta S_{int}^{(1)} + \Delta S_{int}^{(2)}$,
where
\begin{align}
 \label{Sints}
 \Delta S_{int}^{(1)} &= {\hbar\over 2m_0} \int d^2r dt \;
 (\eta \nabla \Phi_V -\phi \nabla \rho_V)\cdot {\bm \dot {\bm R}}_V \nonumber \\
  \Delta S_{int}^{(2)}&= {\hbar\over 2m_0}
  \int d^2r dt \; ( \eta \nabla \phi  - \phi \nabla \eta) \cdot
  {\bm \dot {\bm R}}_V
\end{align}

The form of $\Delta S_{int}^{(1)}$ is simply that of an
overlap integral of the vortex zero mode (\ref{zeromode}) with the
perturbed quasiparticles: they are solutions of the same perturbed
quasiparticle equations of motion and are therefore orthogonal. The effect of $\Delta S_{int}^{(1)}$ is to distort both the zero mode vortex profile and the quasiparticles; for details see ref. \citep{thompsonstamp}. In a properly formulated theory of a vortex and the surrounding quasiparticles, there can exist no linear coupling between them. If we substitute the distortions of the vortex and the quasiparticles back into the full action, we find that this is so. The final result includes

(i) A kinetic term of form
\begin{align}
 \label{SVkin}
\tilde S_{qp}[\tilde \phi_0, \tilde \eta_0] \;=\;  S_V^{kin} \; \; = \;\;  {1\over 2} \int dt \; M_v^0  \dot{\bm R}_V^2
\end{align}
where $\tilde \phi_0, \tilde \eta_0$ are the distorted zero modes. This result defines the ``bare" hydrodynamic mass $M_v^0$, which can be found by calculating the distortion explicitly \cite{effectivemass}. Note the hydrodynamic mass, arising from these distortions, then automatically includes a contribution from the distorted quasiparticle cloud that accompanies it.

(ii) An interaction between the vortex and \underline{pairs} of quasiparticles having momentum-energies $({\bf q}_1, \nu_1), ({\bf q}_2,\nu_2)$; this interaction takes the form \cite{thompsonstamp,frousteycoxstamp}
\begin{equation}
F({\bf q}_1, {\bf q}_2) \;=\; 2 \pi i {\hbar \over m_o^2} {q_1q_2 \over |{\bf q}_1 - {\bf q}_2|^2} \sin \theta_{12} \;\times\; (\delta \left({\bf P'} - {\bf P} + \hbar({\bf q}_1 - {\bf q}_2 \right)  \; \delta \left(E' - E - \hbar(\nu_1 - \nu_2)\right)
 \label{vertex}
\end{equation}
In superfluid $^4$He this interaction has strength $\sim O(\Lambda_0)$; but if we now calculate the correction to the bare hydrodynamic mass coming from this, we find that all the corrections are $\sim O(m_0/M_V^0)$ compared to the bare hydrodynamic mass \cite{frousteycoxstamp}. The dissipative corrections to the real time dynamics of the vortex are similarly small \cite{thompsonstamp}. For this reason, in this paper we have ignored dissipative corrections to the tunneling rates.

It is interesting to ask how we might make these radiative corrections larger. In a
superfluid $^4$He film of thickness $L_z$, we have a vortex mass $M_V \sim m_0 (a_o/L_z)/ ln |R/\xi_o|$, where $R$ is the lengthscale over which the vortex flow field extends,  and $a_o$ is the interparticle spacing. For extremely thin films where $L_z\sim a_0$, $m_0/M_V\sim\mathcal{O}\left(1\right)$ at the beginning of the tunneling process, when $R$  is not much larger than $a_0$. But $m_0/M_V$ falls rapidly as the tunneling proceeds. In any case, once a
vortex has moved away from any boundary, so that $R \gg \xi_o$, the radiative
corrections will be very small, for any film.

\subsubsection{WKB/Instanton Tunneling Rates}
To deal with our problem, we start from an instanton formulation of tunneling \citep{coleman,takagi}, in which tunneling is formulated as the propagation of the system in an inverted potential, arrived at by analytic continuation of the propagator to imaginary time. In our problem we are interested in a real time action of form
\begin{equation}
S_{eff} [Q,\dot{Q}] \;=\; \int dt [\tfrac{1}{2} M(Q) \dot{Q}^2 - V(Q)]
 \label{S-eff}
\end{equation}
where $Q$ is some generalized coordinate, $M(Q)$ a $Q$-dependent mass, and $V(Q)$ some effective potential acting on the mass. The real time propagator for this system is then
\begin{eqnarray}
K(2,1) \;&=&\; \int_1^2 {\cal D}Q \int_1^2 {\cal D}P \; \exp {i \over \hbar} \int_1^2 [P dQ - {\cal H} dt] \nonumber \\
& \rightarrow&\; \int_1^2 {\cal D}Q \; \exp {i \over \hbar} S_{eff} [Q,\dot{Q}]
 \label{K-21}
\end{eqnarray}
where we use a short-hand \citep{jordan} in which $``1"$ and $``2"$ label the initial and final states (whether they be momentum or position states, or something else).

Instanton methods can then be immediately applied to the form for the effective action in (\ref{S-eff}), because even though the mass is $Q$-dependent, the Euclidean effective action takes the standard from $S_{eff}^{Euclid} = i\int d\tau [\tfrac{1}{2} M(Q) \dot{Q}^2 + V(Q)]$, and the usual arguments allowing one to convert this to a WKB tunneling form go through \cite{takagi}.

The key question to be decided here is what are the initial and final field
configurations in the system (ie., what are states ``$1$" and ``$2$"). Physically this
is clear; the initial state ``$1$" is one in which (i) no vortex has yet begun to
nucleate at the protuberance, or (ii) no vortex/anti-vortex pair has yet begun to
nucleate. Thus the initial state is one of smooth superflow around the obstacle or in
the bulk film. In our results for ${\cal T}$ we have simply substracted off the energy
of this background smooth superflow, so this initial vacuum energy is zero. We thus
choose the initial ``entry" state to be this vacuum state. The final ``exit" state
``$2$" (where the vortex or vortex/anti-vortex pair emerge from under the energy
barrier) is then degenerate with the initial state, and the idea is then to fix the
entry and exit points so that ${\cal T}(1) = {\cal T}(2) = 0$.

For our problem, $V(Q) \rightarrow {\cal T}({\bf R}_V)$, where ${\bf R}_V$ is some
vortex coordinate; and $M(Q) \rightarrow M_V({\bf R}_V)$, the vortex effective mass. As noted before, the prefactor $\Omega_0$ is an adjustable parameter
because we do not know the short distance physics in our strongly-correlated system, and
our description is only valid at lengthscales $\gg \xi_o$.

All of this allows us to write the tunneling expression in WKB approximation, for our vortex problem, as
\begin{eqnarray}
\Gamma_{\text{WKB}} \;\; \sim \;\; \Omega_{0}\exp\left\{ -\frac{1}{\hbar}\int_{a}^{b}d
{\bf R}_V\,\left( 2M_V({\bf R}_V) \left| {\cal T}({\bf R}_V) \right|\right)^{\frac{1}{2}}
\right\}
 \label{Gamma-inst2}
 \end{eqnarray}
where we need to fix the entry and exit points $a$ and $b$. The exit point is simply
fixed by the condition ${\cal T}({\bf R}_V = b) = 0$. The entry point is more subtle
because both the effective mass $M_V({\bf R}_V)$ and the potential ${\cal T}({\bf R}_V)$
contain terms $\sim \ln |(R_V - r_o)/\xi_o|$ (for the protuberance problem) or $\sim \ln
|r_{12}/\xi_o|$ (for the vortex/anti-vortex pair nucleation), which are meaningless when
the argument of the logarithm $\rightarrow 0$. To deal with this we fix the entry point
$a$ such that ${\cal T}({\bf R}_V = a) = 0$, and adjust $a$ numerically so this condition is satisfied. One then finds that $a$ to be somewhat greater than $\xi_o$, in all numerical work.

Using eqtn. (\ref{Gamma-inst2}) we then immediately get the results quoted in the main text for the WKB tunneling rates, for the 2 cases which we analyzed there (the semicircular protuberance, and the vortex/anti-vortex pair).

\subsection{Numerical Results: Extrinsic Nucleation at Semi-Circular Bump}

In order to resolve the integrals contained within eqtn. (\ref{Gamma-inst2}), for the tunneling rate, numerical work must be done for both the extrinsic and intrinsic case.

To reduce the complexity of the numerical work, we define the dimensionless tunneling exponent $\gamma\left(\xi_{0},u_{0}\right)$, which contains the relevant integrals needed to be numerically solved. The tunneling rate, written in terms of the dimensionless tunneling exponent is then
\begin{equation}
\Gamma_{\text{WKB}}=\Omega_{0}\exp \left\{ -\gamma\left(\xi_{0},u_{0}\right)\right\}
\label{eq:Dimensionless Tunneling Rate}
\end{equation}

Using eqtn. \eqref{eq:Energy-Bump} for the energy near a wall with a protuberance and equation \eqref{Meff-bump} for the effective mass of the vortex, the extrinsic tunneling rate is given by eqtn. \eqref{eq:Dimensionless Tunneling Rate}, where the tunneling exponent is given from eqtn. $\text{\eqref{eq:Energy-Bump}}$ by
\begin{equation}
\label{eq:Energy-Bump-Unitless}
\gamma\left(\bar{\xi}_{0},u_{0}\right)\;=\; \frac{\rho_{s}\kappa^{2}r_{0}}{2\pi\hbar c_{0}} \int_{x_{a}}^{x_{b}}dx\,\sqrt{\ln\left(\frac{x-1}{\overline{\xi}_{0}}\right)\left[\ln\left(\frac{4x^{2}}{\overline{\xi}_{0}^{2}}\left(\frac{x^{2}-1}{x^{2}+1}\right)^{2}\right)-\cdot u_{0}\left[\left(\frac{x^{2}-1}{x}\right)-\frac{8}{\pi}\tan^{-1}\left(\frac{x}{x^{2}-1}\right)\right]\right]}
\end{equation}
and where we can define a ``critical velocity" $v_c$ at which the tunneling barrier disappears, given by  $v_{c}= \kappa/2\pi R_{0}$, and a dimensionless background velocity $u_{0}=v_{0}/v_{c}$. We also define a dimensionless vortex core radius and dimensionless vortex core distance from the bump, as $\overline{\xi}_{0}= \xi_{0}/r_{0}$
and $x=R_{\text{V}}/r_{0}$ respectively.

\begin{figure}[hbt!]
\begin{centering}
\includegraphics[width=0.5\paperwidth]{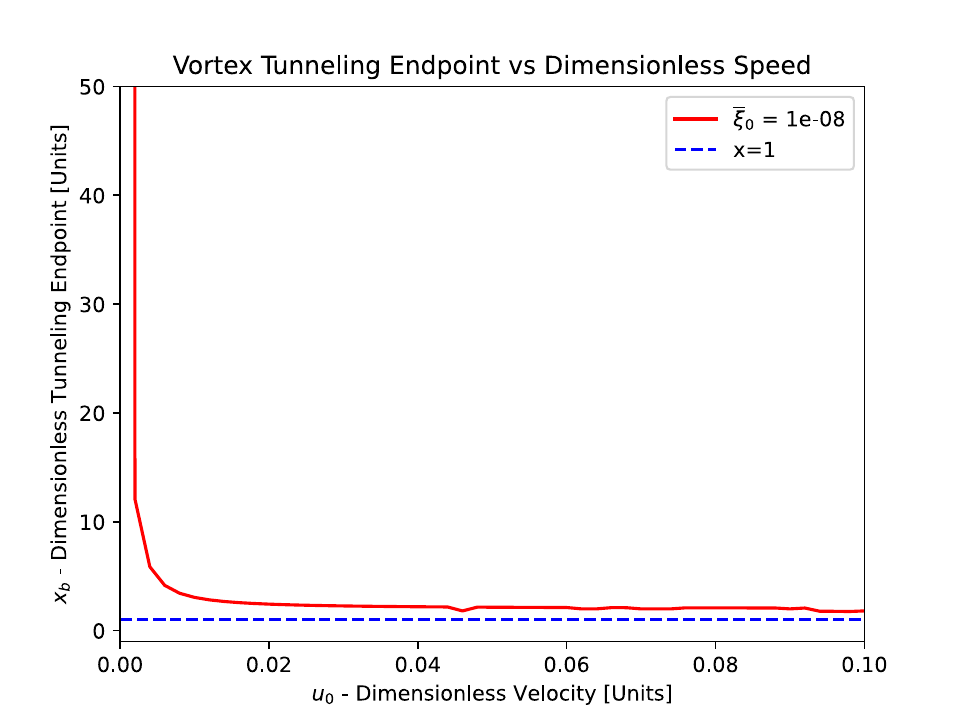}
\par\end{centering}
\caption{\label{fig:bplot}A plot of $x_{b}\left(\overline{\xi},u_{0}\right)$,
the exit point of the tunneling process, which would represent the
location, $x_{b}=r_{0}b$, at which the core of the vortex would
appear after tunneling. Note that for small $u_{0}$, the distance
$x_{b}$ increases without bound, a fact that makes the tunneling
exponent $\gamma$ extremely large, and therefore the tunneling rate
vanishingly small.}
\end{figure}

The first challenge encountered here is that while the initial tunneling point $x_{a}$ always stays near $x=1$, the final tunneling point $x_{b}$ varies quite
a bit as a function of the velocity $u_{0}$. Indeed $x_{b}$ can become unbounded for small speeds, and must be determined numerically. The MINPACK algorithims ``hybrd'' and ``hybrj''{[}14,15{]} are implemented
though the function ``fsolve'' found in the SciPy library. Accounting
for complex roots and the like, $x_{b}$ can be fully determined,
as shown in Figure $\text{\ref{fig:bplot}}$.


A series of $N$ integrals can be numerically determined for $\gamma$, all with different velocities $u_{0}$, to provide a collection of points $\left\{ \gamma\left(u_{0}^{\left(n\right)},\overline{\xi}_{0}^{\left(n\right)}\right)\right\} _{n=1}^{N}$. In order to accomplish this efficiently and accurately, Adaptive Gaussian
Quadrature is used, which is implemented by the Fortran 77 library
QUADPACK \cite{QUADPACK}, and used in a python code.

The objective of all of this integration is to produce a lattice of points $\left\{ \left(u_{0}^{\left(n\right)},\overline{\xi}_{0}^{\left(n\right)}\right)\right\} _{n=1}^{N}$ that can then be interpolated using an appropriate fit function.

This interpolation is done for a lattice of points, which includes varying the vortex core-radius $\overline{\xi}_{0}$. The reason for varying the core-radius in this analysis is to provide a more accurate and more general fit function. The fit function that makes most sense given the asymptotic behaviour
of $\ln\left(\overline{\xi_{0}}\right)$ and $1/u$, was found earlier
in D. Marchand \cite{marchand}; one has
\begin{equation}
\overline{\gamma}_{\text{fit}}\left(\overline{\xi_{0}},u_{0}\right)=\left[A_{0}\ln\left(\overline{\xi_{0}}\right)+A_{1}\right]u^{-\left(B_{0}\overline{\xi_{0}}^{B_{1}}+B_{2}\right)}+\left[C_{0}\ln\left(\overline{\xi_{0}}\right)+C_{1}\right]\label{eq:Fit Function}
\end{equation}

Using the SciPy library in Python to implement the Levenberg-Marquardt
interpolation algorithim \cite{interpolation}, one is able to obtain a variety of parameters for
$\text{\eqref{eq:Fit Function}}$ that will converge to the data generated
numerically. The set of values that worked best can be found, using
sensible initial guesses for the parameters. This parameter family
is shown in Table $\text{\ref{tab:parameterfits}}$ , compared to
the bare vortex mass numerical work described by D.Marchand \cite{marchand}.


\begin{table}[h]
\begin{centering}
\begin{tabular}{|c|c|c|}
\hline
Parameter & $\begin{array}{c}
\overline{\xi}_{0}\in\left(10^{-8},0.1\right)\\
u_{0}\in\left(10^{-10},20\right)
\end{array}$ & Bare Mass\tabularnewline
\hline
\hline
$A_{0}$ & -94.72 & -12.0\tabularnewline
\hline
$A_{1}$ & 534 & -18\tabularnewline
\hline
$B_{0}$ & 0.27  & 0.27\tabularnewline
\hline
$B_{1}$ & 0.21 & 0.21\tabularnewline
\hline
$B_{2}$ & 1.04 & 1.041\tabularnewline
\hline
$C_{0}$ & 37.63 & 0.171\tabularnewline
\hline
$C_{1}$ & -350 & -1.041\tabularnewline
\hline
\end{tabular}
\par\end{centering}
\caption{\label{tab:parameterfits}Fit parameters for the fit function $\text{\eqref{eq:Fit Function}}$,
determined using the Levenberg-Marquardt alorithim using a lattice
of points $\left\{ \gamma^{\left(n\right)},u_{0}^{\left(n\right)},\overline{\xi}_{0}^{\left(n\right)}\right\} _{n=1}^{N}$.
The first column is for the case where the dynamic effective mass
is taken into account, and the second column is the fit parameters
for when only the bare mass is taken into account. }
\end{table}


With this fit, one should be able to produce a plot of the dimensionless tunelling coefficient as a function of $\overline{\xi}_{0}$ and $u_{0}$.

\subsection{Numerical Results: Intrinsic Nucleation of Vortex/Anti-Vortex Pairs}
In the case of intrinsic nucleation, the tunneling rate must also be determined numerically using the same numerical methods described in the previous section. The dimensionless tunneling rate, using the expression for energy from equation $\text{\eqref{eq:Energy - Vortex/Antivortex Pair}}$ and the effective vortex mass from equation $\text{\eqref{Mv-V-aV}}$, is just
\begin{equation}
\Gamma_{\text{WKB}}=\Omega_{0}\exp\left\{ -\frac{\sqrt{2}\xi_{0}\rho_{s}\kappa^{2}}{2\pi\hbar c_{0}}\int_{x_{a}}^{x_{b}}\sqrt{\ln\left|x\right|\cdot\left(\ln\left|x\right|-u_{0}x\right)}dr\,\right\}
 \label{eq:Tunneling Rate - Intrinsic}
\end{equation}
where $x=q/\xi_{0}$ is the dimensionless distance between the vortex/anti-vortex pair, and $u_{0} = v_o/v_c$ is again the dimensionless superflow velocity; and where $x_{a}=a/\xi_{0}$ and $x_{b}=b/\xi_{0}$ are the entrance and exit tunneling endpoints.

This tunneling exponent is the product of a dimensionless integral over $x$ and a dimensionless prefactor $\sqrt{2}\rho_{s}\kappa^{2}\xi_{0}/2\pi\hbar c_{0}$. We can rewrite this prefactor as 
\begin{equation}
\frac{\sqrt{2}\xi_{0}\rho_{s}\kappa^{2}}{2\pi\hbar c_{0}} \;\; = \;\; 0.88 {L_z \over a_o}
 \label{exp-prefactor}
\end{equation}
where we have rewritten $\rho_s = (L_z/a_o) \rho_s^{single}$, in which $\rho_s^{single}$ is the superfluid density for a single layer of superfluid, of trhickness $a_o$. We assume here that $a_o = 3.4 $ Angstroms, with a bulk 3-d superfluid density $\sim 172 \; kgm^{-3}$, appropriate to the superfluid on the melting curve. 

To get the final tunneling exponent we then have to multiply this by the second (also dimensionless) part of the exponent, the integral over $x$. It is possible to find the relevant zeroes of the integrand analytically. To find $x_{a}$ and $x_{b}$ one must find the roots
of $\mathcal{T}\left(x\right)$ in equation $\text{\eqref{eq:Energy - Vortex/Antivortex Pair}}$. To achieve this, we start with $u_{0}x=\ln\left(x\right)$ and take the exponential of both sides to get
\begin{equation}
x=e^{u_{0}x}
\end{equation}

This is a particular form of the expression $x=a+be^{cx}$, which
has the following solutions \cite{mezo},
\begin{equation}
x=a-\frac{1}{c}\text{LambertW}\left(n,-bce^{ac}\right)
\end{equation}
where $n$ corresponds to different branch cuts of this special function.
In our case, $c=u_{0}$, $a=0$, and so our result reduces to the solution
\begin{equation}
x=-\frac{1}{u_{0}}\text{LambertW}\left(n,-u_{0}\right)
\end{equation}

Only two of these branch cuts are useful, viz., $n=0$ and
$n=-1$, which correspond to $x_{a}$ and $x_{b}$ respectively,
\begin{align}
x_{b}\left(u_{0}\right) & =-\frac{\text{LambertW}\left(-1,-u_{0}\right)}{u_{0}}\nonumber \\
x_{a}\left(u_{0}\right) & =-\frac{\text{LambertW}\left(0,-u_{0}\right)}{u_{0}}\label{eq:Intrinsic-Upperandlowerbounds}
\end{align}


\begin{figure}[hbt!]
\begin{centering}
\includegraphics[width=0.41\paperwidth]{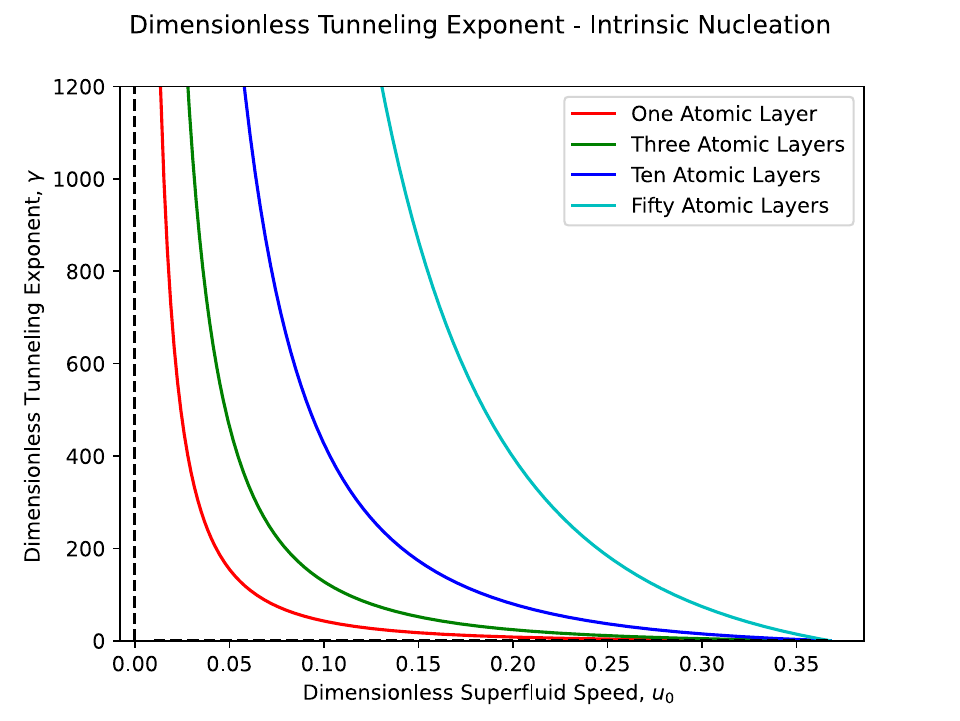}\hfill
\includegraphics[width=0.41\paperwidth]{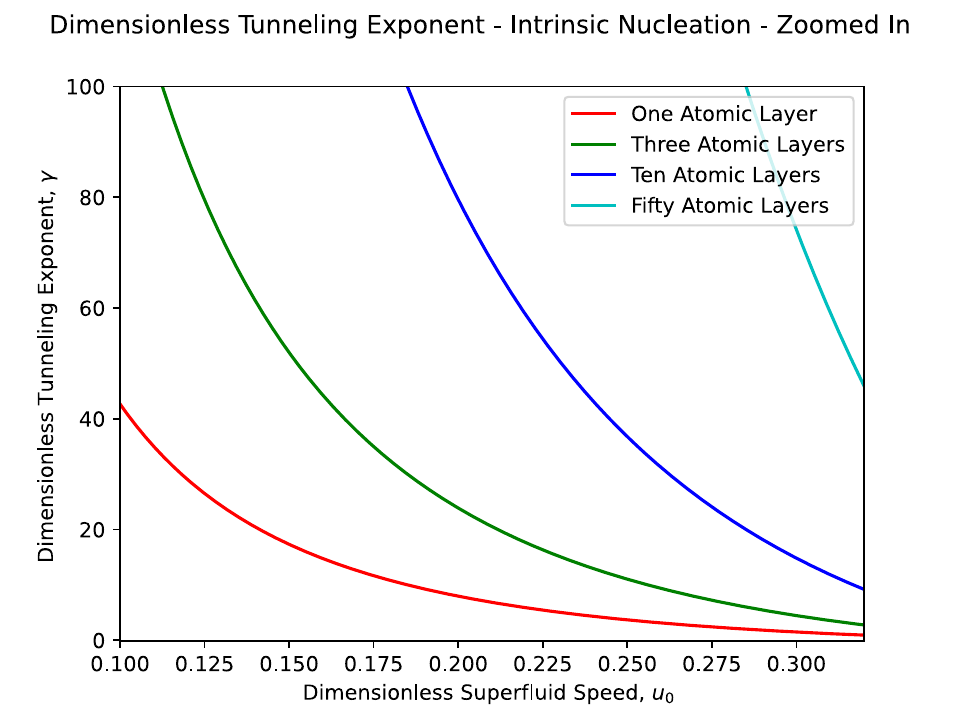}
\par\end{centering}
\caption{\label{fig:IntrinsicTunnelingExponent}Two plots of the tunneling exponent $\gamma\left(u_{0}\right)$ as a function of the background superfluid velocity $u_{0}$, for different atomic layers. The figure on the right is zoomed in to highlight where the tunneling rates begin to rapidly increase; namely where the tunneling exponents begin to vanish. The number of atomic layers determines the 2D density through equation \eqref{exp-prefactor}}
\end{figure}


The dimensionless tunneling exponent, which is the integral found in equation \eqref{exp-prefactor},
using the expressions for the upper and lower bounds, eqtn. $\text{\eqref{eq:Intrinsic-Upperandlowerbounds}}$,
is a function of $u_{0}$ only and has a simple fit function, which has been found numerically,
\begin{equation}
\tilde{\gamma}_{\text{fit}}\left(u_{0}\right)=\frac{a}{\left(u_{0}+b\right)^{c}}+d\label{eq:Dimensionless Tunneling Exponent-Intrinsic- FitFunction}
\end{equation}
where the fit parameters are, $a=6.84916255\times 10^{-1}$, $b=4.47057519\times10^{-3}$ , $c= 1.89427958$ and $d=-10.0000000$.
As with the extrinsic case, the probability of nucleation from tunneling
increases as the superfluid flow speed $u_{0}$ increases.

We can now see what the results look like. Let's first look at how the tunneling exponent $\gamma(u_o)$ varies with $u_0$; this is shown for a number of different superfluid film thicknesses. This is shown in Fig \ref{fig:IntrinsicTunnelingExponent}


\begin{figure}[hbt!]
\begin{centering}
\includegraphics[width=0.41\paperwidth]{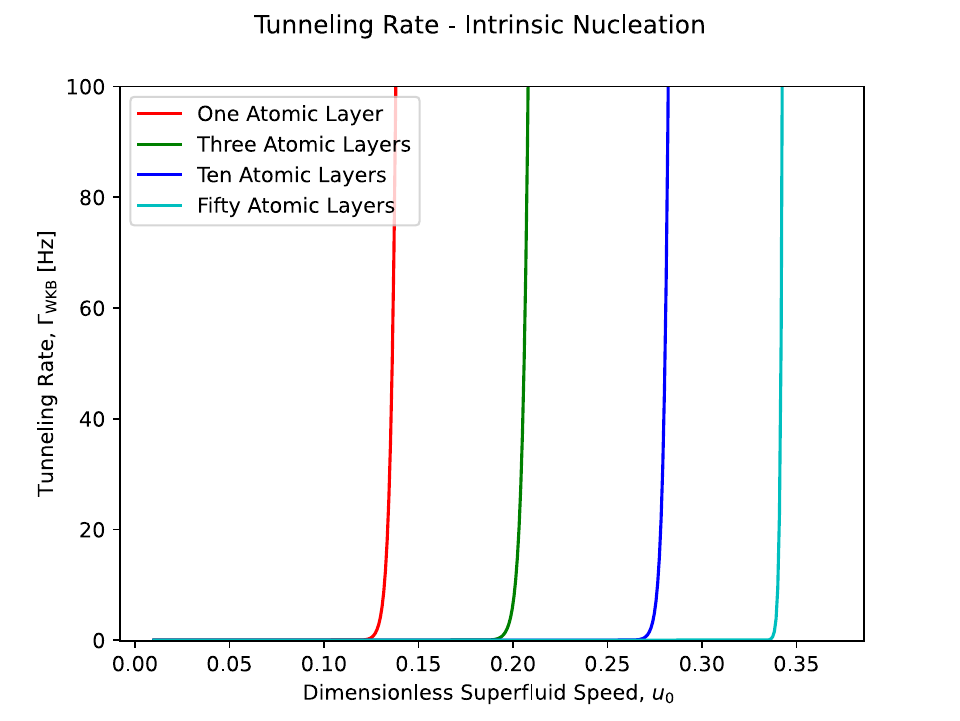}\hfill
\includegraphics[width=0.41\paperwidth]{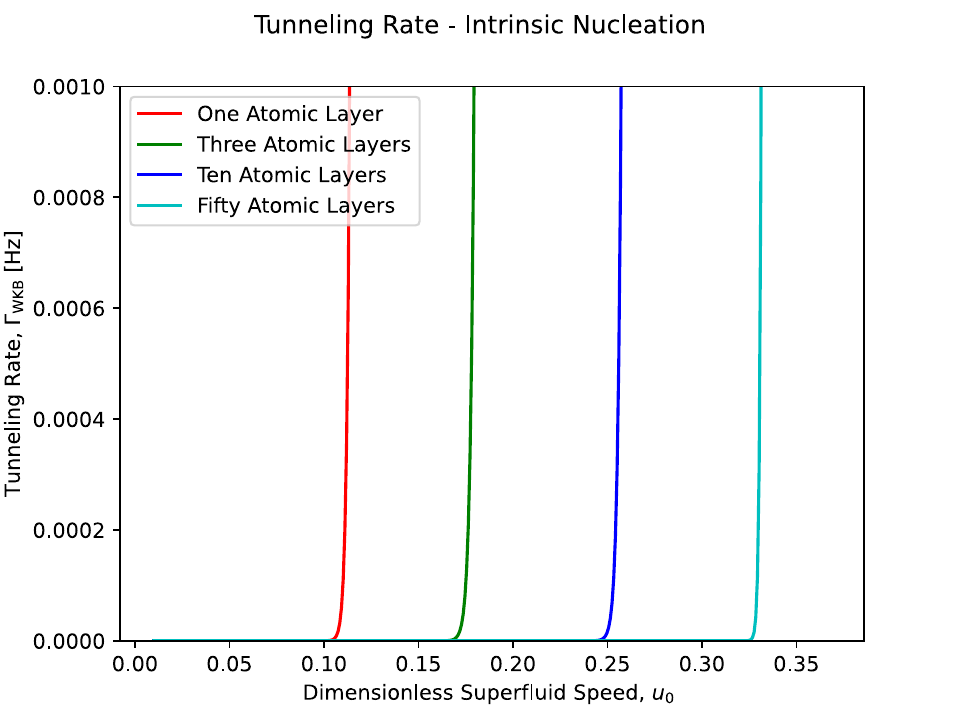}
\par\end{centering}
\caption{\label{fig:IntrinsicTunnelingRateLinear}Two plots of the tunneling rate \eqref{exp-prefactor} for the Intrinsic Nucleation of a vortex/anti-vortex pair, for various number of atomic layers and where $\Omega_{0}\sim $ 150GHz. In these plots it is easy to see at which superfluid speed the vortex/anti-vortex pair beings to rapidly nucleate.}
\end{figure}


We then look at the tunneling rates computed from this, again for different film thicknesses. We see that this rate varies extremely rapidly in the crossover regime. In experiments this will look almost like a sudden transition (see Fig.\ref{fig:IntrinsicTunnelingRateLinear}). That this is not the case is seen if we change the time window on which we look at the transtion.


\begin{figure}[hbt!]
\begin{centering}
\includegraphics[width=0.5\paperwidth]{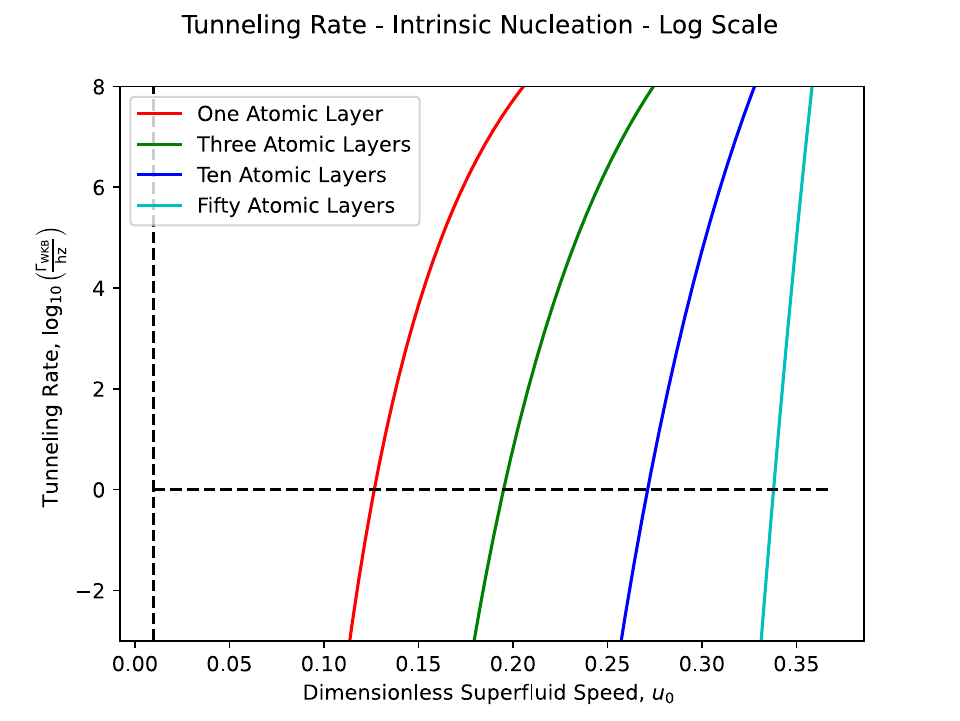}
\par\end{centering}
\caption{\label{fig:IntrinsicTunnelingRateLog}Two plots of the tunneling exponent $\gamma\left(u_{0}\right)$ as a function of the background superfluid velocity $u_{0}$, for different atomic layers. The figure on the right is zoomed in to highlight where the tunneling rates begin to rapidly increase; namely where the tunneling exponents begin to vanish. The number of atomic layers determines the 2D density through equation \eqref{exp-prefactor}}
\end{figure}


This last point is reinforced in Fig. \ref{fig:IntrinsicTunnelingRateLog}, where we now plot the logarithm of the tunneling rate for different film thicknesses as a function of $u_0$. The nature of the crossover is very obvious from these plots.
